\documentclass[fleqn,10pt]{wlscirep}
\usepackage[utf8]{inputenc}
\usepackage[T1]{fontenc}
\usepackage{indentfirst}
\usepackage{todonotes}
\usepackage{hyperref}
\usepackage{booktabs,longtable}
\usepackage{pdflscape}
\usepackage{floatrow}
\usepackage{parskip}
\usepackage{longtable}
\usepackage{soul}
\usepackage{adjustbox}
\DeclareFloatFont{tiny}{\small}
\usepackage{xcolor}
\usepackage{array}
\newcolumntype{P}[1]{>{\raggedright\let\newline\\\arraybackslash\hspace{0pt}}p{#1}}
\usepackage{amsmath}

\title{COVID-19 Activity Risk Calculator as a Gamified Public Health Intervention Tool}

\author[1*]{Shreyasvi Natraj}
\author[3]{Malhar Bhide}
\author[3]{Nathan Yap}
\author[4]{Meng Liu}
\author[5]{Agrima Seth}
\author[6]{Jonathan Berman}
\author[2,3*]{Christin Glorioso}
\affil[1]{Department of Neurosciences, Faculty of Medicine, University of Geneva, Geneva, Switzerland}
\affil[2]{Department of Anatomy, University of California, San Francisco, CA, USA}
\affil[3]{Academics for the Future of Science Inc., Cambridge, MA, USA}
\affil[4]{Department of Industrial and Manufacturing Engineering, Penn State University, PA, USA}
\affil[5]{School of Information, University of Michigan, Ann Arbor, MI, USA}
\affil[6]{Department of Basic Science, New York Institute of Technology College of Osteopathic Medicine at Arkansas State University, AR, USA}
\affil[*]{shreyasvi.natraj@unige.ch}
\affil[*]{gloriosoca@gmail.com}


\begin{abstract}

The Coronavirus disease 2019 (COVID-19) pandemic, caused by the virus severe acute respiratory syndrome coronavirus 2 (SARS-CoV-2), has impacted over 200 countries leading to hospitalizations and deaths of millions of people. Public health interventions, such as risk estimators, can reduce the spread of pandemics and epidemics through influencing behavior, which impacts risk of exposure and infection.
Current publicly available COVID-19 risk estimation tools have had variable effectiveness during the pandemic due to their dependency on rapidly evolving factors such as community transmission levels and variants. There has also been confusion surrounding certain personal protective strategies such as risk reduction by mask-wearing and vaccination. In order to create a simple easy-to-use tool for estimating different individual risks associated with carrying out daily-life activity, we developed COVID-19 Activity Risk Calculator (CovARC). CovARC is a gamified public health intervention as users can "play with" how different risks associated with COVID-19 can change depending on several different factors when carrying out routine daily activities. Empowering the public to make informed, data-driven decisions about safely engaging in activities may help to reduce COVID-19 levels in the community. In this study, we demonstrate a streamlined, scalable and accurate COVID-19 risk calculation system. Our study also demonstrates the quantitative impact of vaccination and mask-wearing during periods of high case counts. Validation of this impact could inform and support policy decisions regarding case thresholds for mask mandates, and other public health interventions. 

\end{abstract}

\begin{document}

\flushbottom

\maketitle
\thispagestyle{empty}

\section*{Introduction}

The risk levels for gatherings and activities during the COVID-19 pandemic have fluctuated significantly since its onset\cite{Martinez-Perez2020-va,Subramanian2022}. These fluctuations have been driven by variation in the determinants of infection, hospitalization, and death risk, which include community transmission levels, size of gatherings, social distancing, vaccination status, type of vaccine, the dose of vaccine, face mask usage, air filtration, circulating SARS-CoV-2 variants, health conditions, age, gender, previous infection, and time since previous infection or vaccination (due to waning immunity)\cite{roy_ghosh}. The risk of becoming infected with SARS-CoV-2 in the environments most conducive to spread (indoor and crowded) are significantly higher at peak community case levels than at lower community case levels\cite{10.1371/journal.pmed.1003871,doi:10.1177/1178633721991260,E_Vincenzo2021-vu}. Due to the rapidly evolving nature of these risk levels and the number of determinants involved\cite{Haas2020-fd}, it has been challenging for the public to maintain a clear understanding of what their risk is for various activities over time. As a result, some non-scientist citizens have created tools\cite{microcovidproject} to try to assess risk for their households, and publications such as the New York Times\cite{nytimesarticle} have attempted to create risk assessment algorithms. However, these manual attempts can be complicated to use and lacking in enough determinants to accurately estimate risk, resulting in confusion over cost-benefit surrounding harm reduction strategies such as the use of face masks, restrictions on gathering size, and vaccination, creating an "infodemic"\cite{infodemic}.

Research teams have created various tools for individuals to evaluate their risk\cite{evaluation_2022} (see \href{tab:comparison}{Table \ref{tab:comparison}}), each taking a somewhat different approach. A simple approach taken by many Dashboards, including the US Center for Disease Control (CDC) is to report risk levels on a county-wide basis simply by community transmission levels \cite{coviddatatracker}. While this is a simple to understand and a useful metric, it doesn't address differences in risk by individual risk factors such as age and health condition, differences in activities such as number of attendees, or differences in precautions such as mask-wearing or vaccination. All of these factors are important and can change risk drastically. 

\begin{table}
\centering
\begin{adjustbox}{width=\columnwidth,center}
\begin{tabular}{|l|l|l|l|l|l|l|l|l|l|l|l|l|} 
\hline
\textbf{Risk Calculator} & \textbf{Is} & \multicolumn{8}{l|}{\textbf{Considers}} & \multicolumn{3}{l|}{\textbf{Calculates the risk of}} \\ 
\cline{2-13}
 & Fast & Vaccination & \begin{tabular}[c]{@{}l@{}}Health\\Conditions\end{tabular} & \begin{tabular}[c]{@{}l@{}}Many\\Locations\end{tabular} & \begin{tabular}[c]{@{}l@{}}Mask \\Type\end{tabular} & \begin{tabular}[c]{@{}l@{}}Number of\\People\end{tabular} & \begin{tabular}[c]{@{}l@{}}Indoor or\\Outdoor\end{tabular} & Variants & \begin{tabular}[c]{@{}l@{}}Specific\\Activities\end{tabular} & Infection & Hospitalization & Death \\ 
\hline
\begin{tabular}[c]{@{}l@{}}CovARC (Our Calculator)\end{tabular} & X & X & X & X & X & X & X & ~ ~~X & & X & ~ ~ ~ ~X & X \\ 
\hline
QCovid\cite{qcovid} & X & & X & & & & & & & & ~ ~ ~ ~ ~X & \\ 
\hline
ASIMI Model\cite{ASIMI} & X & & & & X & & X & & & X & & \\ 
\hline
\begin{tabular}[c]{@{}l@{}}19 an Me\\(Princeton Model)\cite{19andme}\end{tabular} & & X & X & & & & & & X & X & & \\ 
\hline
MyCOVIDRisk\cite{mycovidrisk} & X & & & & X & X & X & & X & X & & \\ 
\hline
\begin{tabular}[c]{@{}l@{}}COVID-19\\Assessment Tool\\(GA Tech Model)\cite{covidgatech}\end{tabular} & X & & & X & & & & & & X & & \\ 
\hline
Covidtracker.fr\cite{covidtracker} & X & & & & & & & & & X & & \\ 
\hline
\begin{tabular}[c]{@{}l@{}}Max Planck Institute\\COVID Risk Calculator\cite{maxplancktracker}\end{tabular} & & & & & X & & X & X & & X & & \\ 
\hline
\begin{tabular}[c]{@{}l@{}}COVID-19 Indoor Safety\\Guide (MIT Model)\cite{mitriskcalculator}\end{tabular} & X & & & & X & & X & X & & X & & \\ 
\hline
\begin{tabular}[c]{@{}l@{}}COVID-19 Risk Calculator\\(Northwestern University\\Model)\cite{northwestedu}\end{tabular} & X & & X & & & & & & & & & X \\ 
\hline
microCOVID Project\cite{microcovidproject} & & X & & X & X & X & X & & & X & & \\
\hline
\end{tabular}
\caption{\small{This table provides a comparison between different risk calculators on the basis of the time each user needs in order to carry out the risk estimation process (Fast or not), the factors that each of the risk calculator considers when carrying out risk calculation (whether it considers vaccination, usage of mask, past health conditions, presence of COVID-19 variants, number of people the user is going to be in contact with and if the specific activity the user is going to be performing is indoors or outdoors). We also include number of locations around the world that the calculator works for and if they take into account specific activities (such as going to supermarket, football practice etc.). We then check if there the currently existing risk calculators estimate overall risk or also provide risk of infection, hospitalization and death separetely. We then identify the drawbacks and areas where the currently existing risk calculators are lacking in order to create our risk calculator which is fast, takes into account a majority of factors when estimating risk and estimates different levels of risk (infection, hospitalization and death).}}
\label{tab:comparison}
\end{adjustbox}
\end{table}

The approach factoring in the smallest number of determinants is the  COVID-19 Event Risk Assessment Planning Tool, a web-based tool \cite{covidgatech}, developed by scientists at the Georgia Institute of Technology. The tool estimates the probability that a person will encounter someone infected with SARS-CoV-2 at a particular gathering based on the group's size and the event's location (factoring in under-reporting adjusted community transmission levels). The tool works for the US only, at the county level and does not consider the individual's health risk factors, vaccination status, variants, mask usage, or indoor v. outdoor activities. It estimates the chances that someone at a gathering of a user-set size will be actively infected with SARS-CoV-2.  

The 19 and Me calculator \cite{19andme}, developed by Mathematica, a policy-research company in Princeton, New Jersey, draws on demographic and health information and user behaviors such as hand washing and masks used to determine the relative risk of exposure, infection, and severe illness for the individual's behavior every week. It does not account for variants and only works for the US and Belgium. 

In December 2020, a team led by biostatistician Nilanjan Chatterjee at Johns Hopkins University in Baltimore, Maryland, released the COVID-19 Mortality Risk Calculator \cite{jin2020assessment}, which estimates an individual's relative risk of death from COVID-19 during an activity based on their location, pre-existing conditions, and general health status. It also reports the estimated risk of death in a user's area in the next two weeks. Activities, personal precautions, vaccination, and variants are not factored in.  

Another approach for risk calculator called MyCOVIDRisk \cite{mycovidrisk} takes a more situational approach, estimating the risks associated with specific errands or recreational activities. The estimate is based on the location, duration, and the number of masked or unmasked people attending. This can help users avoid activities that are likely to be high risk in a specific pandemic hotspot, such as spending an hour or more at an indoor gym, favoring safer alternatives — a masked meet-up in the park, for instance, \cite{Rowe2020.12.30.20249058}. It does not report risk numerically, instead opting for a low-very large scale, which does not consider an individual's risk tolerance threshold. It also does not account for variants.

We build upon the comprehensiveness of these tools with more factors taken into account and more geographical locations. To our knowledge, this estimation method is the only one that considers variants and vaccine coverage and works in almost every country. The app is lightweight and can be used on low bandwidth internet, which is an important factor in many countries. We aim to aid people worldwide in making informed decisions about how to do activities more safely during a global Pandemic.

\section*{Results}
\begin{figure}[hbt!]
\centering
\includegraphics[width=\textwidth]{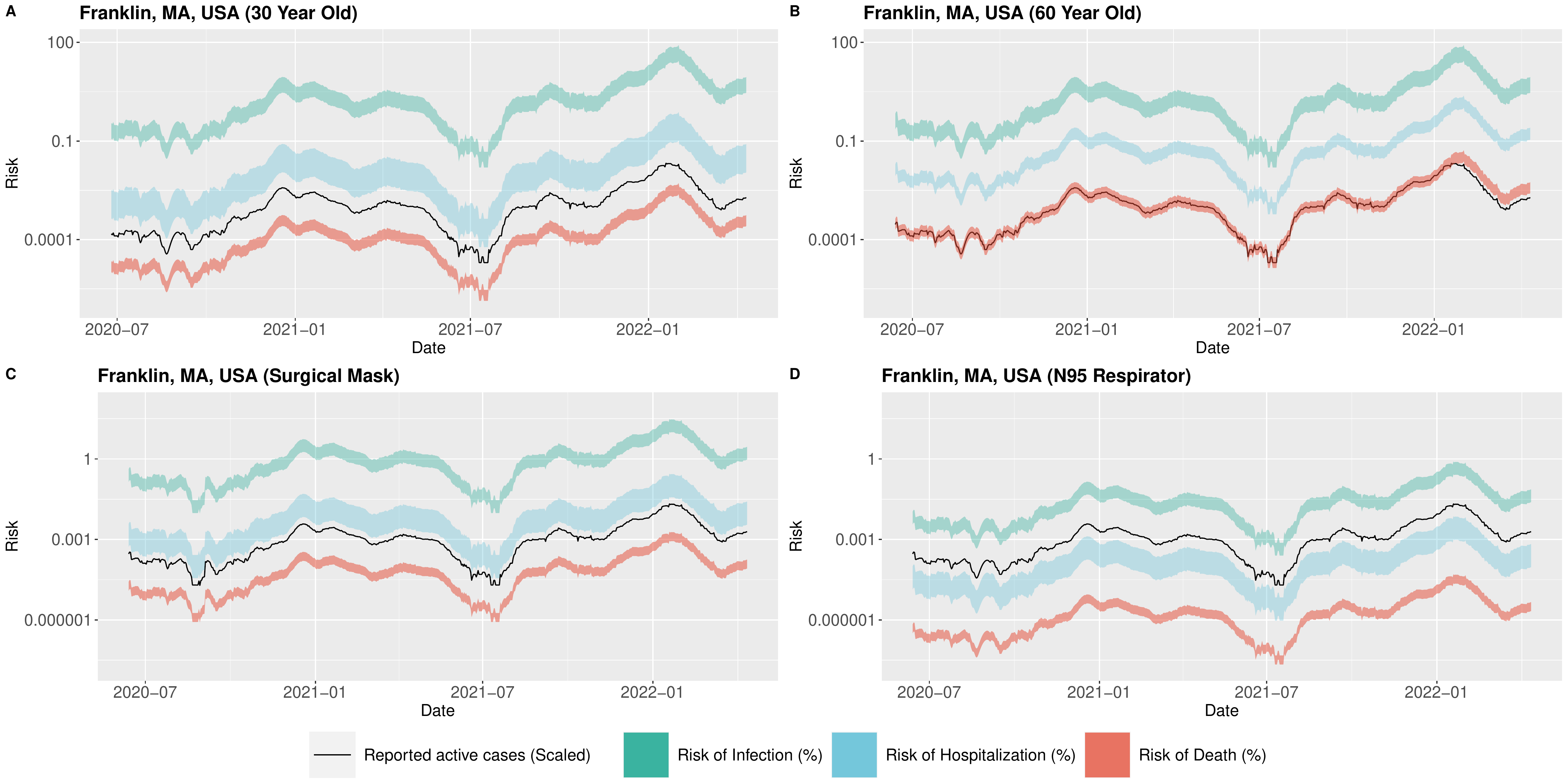}
\includegraphics[width=\textwidth]{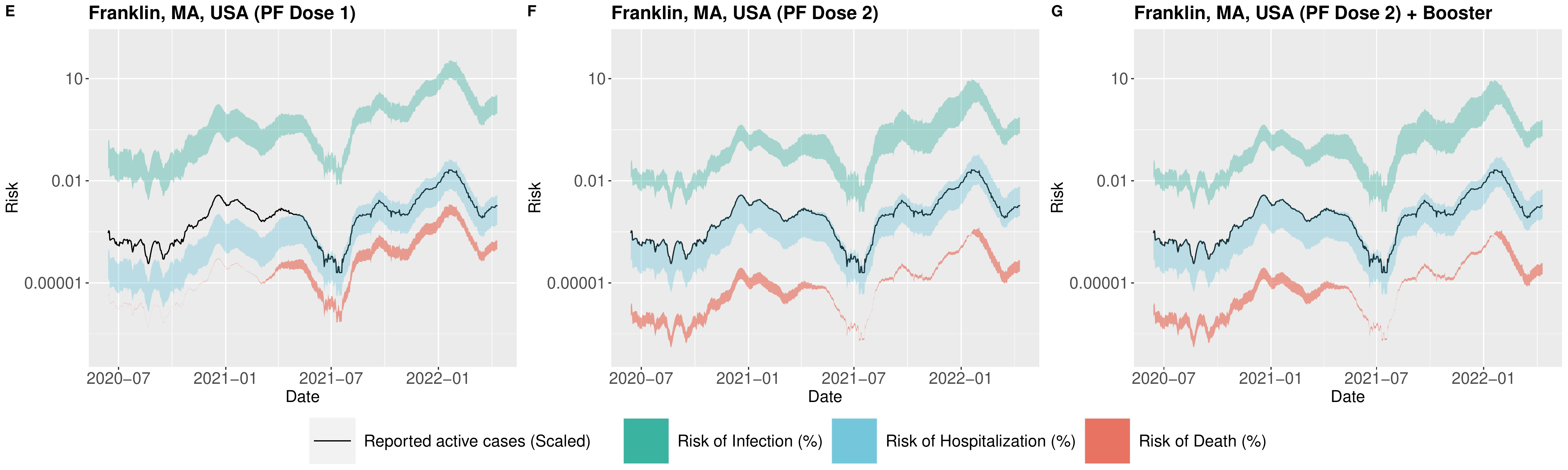}
\caption{\small{In the following figures, we use the location of Franklin, Massachusetts, United States of America and calculated the range of risks of infection, hospitalization and death for a \textbf{A.} 30-year-old male with no chronic illness, no mask and no vaccination, 10 people passed outdoors and 5 people passed indoors during the activity, a \textbf{B.} 60-year-old male with no chronic illness, no mask and no vaccination, 10 people passed outdoors and 5 people passed indoors during the activity. a \textbf{C.} 30-year-old male with no chronic illness, surgical mask and no vaccination when 10 people are passed outdoors and 5 people passed indoors during the activity, a \textbf{D.} 30-year-old male with no chronic illness, N95 respirator mask and no vaccination when 10 people are passed outdoors and 5 people passed indoors during the activity, a \textbf{E.} 30-year-old male with no chronic illness, no mask and Dose 1 of Pfizer vaccination when 10 people are passed outdoors and 5 people indoors during the activity, a \textbf{F.} 30-year-old male with no chronic illness, no mask and Dose 2 of Pfizer vaccination when 10 people are passed outdoors and 5 people indoors during the activity and a \textbf{G.} 30-year-old male with no chronic illness, no mask and Dose 2 with a booster dose of Pfizer vaccination when 10 people are passed outdoors and 5 people indoors during the activity.}}
\label{fig:graph_1}
\end{figure}

We implement here a simple COVID-19 Activity Risk Calculator (CovARC). We extract the number of confirmed cases by using the Johns Hopkins dataset\cite{CSSEGISandData, Zelenova2021-lt} and cross-validate the number of confirmed cases using the Facebook surveys \cite{surveys} dataset in order to obtain the upper and lower limits of the number of confirmed cases. We subtract confirmed cases from the preceding day's number of confirmed cases in order to identify the number of active cases. These active cases are then taken into account to create a 14-day aggregate. We then divide the 14-day aggregate by the population of the city/state in order to identify the density of COVID-19 cases. We use this as the actual density of active cases which is then taken by the risk calculator in order to estimate the risk. Apart from the number of active cases, we take into account the presence of alpha, beta, gamma, delta and omicron variants using the variants dataset\cite{cdcvariants} to identify the influence of variants on the risk of infection. The variants data is coupled with the vaccine efficacy data (see \href{tab:mask_ffe_eff}{Table \ref{tab:mask_ffe_eff} B.}) in order to obtain the final influence of variants and vaccines on the overall simulated range of risk of infection (see \href{fig:graph_1}{Figure \ref{fig:graph_1} D., E., F.}). We also took into account the influence of mask wearing (see \href{tab:mask_ffe_eff}{Table \ref{tab:mask_ffe_eff}} A.) on the overall simulated range of risk of infection from COVID-19 (see (\href{fig:graph_1}{Figure \ref{fig:graph_1} C., D.}). Since there are several types of risk associated with COVID-19 (infection, hospitalization, and death), we included inputs that change these risks differentially, including COVID-19 variants, and health and demographic factors of the user including, age, sex, and chronic illness (see \href{tab:hosp_death}{Table \ref{tab:hosp_death}}) in order to obtain the simulated range of risks of hospitalization and death for different scenarios (see \href{fig:graph_1}{Figure \ref{fig:graph_1} A., B.}).

Using several different datasets and references we obtained the results using our risk calculation system over time showcasing the first outbreak of COVID-19 followed by alpha, beta, gamma, and delta variants outbreak and the omicron variant outbreak. Using our risk calculator, we demonstrate that the risk level increases with a high number of active cases, older age, less vaccination, and lower quality or no face mask usage (see \href{fig:graph_1}{Figure \ref{fig:graph_1}}). We also observed that the risk also increases when carrying out indoor activities, more density of COVID-19 cases, vaccine type, previous chronic illnesses, male sex, and presence of variants of concern (see 
 \href{tab:hosp_death}{Table \ref{tab:hosp_death}}). 

We additionally implemented the code in the form of a streamlined web application that inputs user supplied variables and generates their simulated risk scores (see \href{fig:webpage}{Figure \ref{fig:webpage}}) to create a tool that can be used by anyone anywhere in the world to get a risk score related to COVID-19 when they are carrying out a daily life activity. A simplified flow diagram that outlines the risk calculator's algorithm can be found in \href{fig:wireframe}{Figure \ref{fig:wireframe}}.

To illustrate how risk changes temporally and with different variables, we looped over time from June 2020 until April 2022 under various conditions and plotted the results. The different test scenarios that were performed can be seen as follows.

\subsection*{Change in Risk of Infection, Hospitalization and Death with Community Transmission Levels and Age}
In \href{fig:graph_1}{Figure \ref{fig:graph_1} A.}, we first take the scenario of a 30-year-old male with no chronic health conditions located in Franklin, Massachusetts, USA. We set the conditions to be that the individual is not vaccinated and is not wearing a mask. We set the activity of this person to involve close contact (within 6 feet) with five people indoors and ten people outdoors. We then observe the simulated range of risks of infection, hospitalization, and death over time. There was a close correlation between the change in the number of active cases and  the simulated range of risks of infection, hospitalization, and death. 

The three peaks in the plot highlight the three COVID-19 outbreaks\cite{Wu2020-ju} that took place in the world, the original outbreak, the outbreak of the alpha, beta, gamma, and delta variants, as well as the third outbreak of omicron. We then used the same inputs but altered the age of the individual to 70 years old and observed the same simulated range of risk of infection but a much-elevated simulated range of risk of hospitalization and death (see \href{fig:graph_1}{Figure \ref{fig:graph_1} B.}) indicating that the range of risks of hospitalization and death are higher for an older individual compared to an individual who is younger.

\subsection*{Change in Risk of Infection with Vaccination and Dosage}
We  explored the extent that vaccination and boosting with various vaccines decreases the risk of COVID-19. In order to do so we took the same scenario of a 30-year-old male with no chronic illness located in Franklin, Massachusetts, USA who is not vaccinated and compared it to if he had the first dose of the Pfizer Vaccine, the second dose of Pfizer Vaccine, and the second dose of the Pfizer Vaccine plus the Moderna/Pfizer Booster Vaccine. Using these three scenarios, we again calculated simulated risk for dates spanning the length of the COVID-19 Pandemic and obtained the results (see \href{fig:graph_1}{Figure \ref{fig:graph_1} E., F., G.}).

We observed a reduced risk of infection upon vaccination with the first dose of the Pfizer vaccine and further risk reduction with dose 2 and booster dosage. We also observed that the risk of infection with omicron variants (observed from the third peak) is not significantly decreased by the dosage of the vaccination compared to other variants (observed in the second peak) indicating that precautions are still needed to be followed to ensure that there is less chances of infection from omicron variant.

\subsection*{Change in Risk of Infection with Mask Type}
We examined risk stratification by the usage of face masks. We experimented by taking 3 different scenarios using the first persona of an unvaccinated 30-year-old male with no past chronic illness in Franklin, Massachusetts, USA. We compared the original scenario where the individual was not wearing a mask to two other scenarios where he is wearing a 2-layer woven nylon mask with a nose bridge and one where he is wearing an N95 respirator\cite{maskffearticle}. 

We observed that there is a significant reduction in risk of infection upon wearing a nylon mask with a nose bridge and a very high reduction in risk of infection when wearing the N95 respiration which was in line with several past studies conducted \cite{useofmaskscdcarticle}. Moreover, depending upon the fitted filtration efficacy of the mask, there was a significant reduction in the risk of COVID-19 for both outdoor and indoor activities (\href{fig:graph_1}{Figure \ref{fig:graph_1} C., D.}). This study also helped us to understand the extent of the decrease in risk of COVID-19 by just usage of masks and indicated that masks played a more important role compared to vaccination in the prevention of risk of COVID-19 when there is an outbreak.

We further carried out another secondary test by taking the location of Delhi, India to estimate different risks associated with COVID-19 as well as a reduction in risk due to different dosages of vaccination and usage of masks to check the robustness of our risk calculator for different locations across the world (see \href{fig:graph_2}{Supplementary Figure S1 A., B., C., D., E., F.})

\subsection*{Ease of Usage and Streamlined Estimation of COVID-19 Risk}
After carrying out these studies, we proceeded to make a user interface for CovARC so that it could be implemented as a usable tool by the general public\cite{riskcalculatorlink}. Using RShiny\cite{rshiny}, we deployed a web application that could be used as a way for any user to estimate different risks with carrying out a specific daily life activity by providing input of their specific case. 

By identifying the significant inputs required for calculating several risk factors pertaining to COVID-19 and minimizing the user inputs, we streamlined the risk calculation process at low bandwidths (see \href{fig:webpage}{Figure \ref{fig:webpage}}). Our goal was to implement our study as a tool that could be used on a day-to-day basis. Furthermore, the tool could also enable the user to get acquainted with the range of risks of infection, hospitalization, and death and identify ways in which the risks could be reduced. 

\section*{Discussion}
We demonstrate that CovARC is an accurate and valuable tool for the public and policymakers. The risk estimation system is more comprehensive and simpler than existing alternatives. Due to its streamlined interface, the user is enabled to get acquainted with the range of risks of infection, hospitalization, and death in just a few minutes. This is complemented with comprehensive inputs including age, gender, health conditions, active cases, vaccination (type and dose), use of any face mask, and the number of people in close contact outdoors or indoors. 

We make sure to consider all the important factors when evaluating various risks associated with COVID-19. However, due to the inherent uncertainty in accurately estimating risks, we provide a range of potential risks rather than a single value. Additionally, we assume that individuals that the user encounters are unvaccinated and not wearing masks, as we are unsure about the extent of COVID-19 precautions in a given area. This assumption could slightly inflate the risk in situations where there is high adherence to mask-wearing and vaccination. In future risk assessments for private gatherings or venues with proof of vaccination requirements, we suggest incorporating the behavior of others. We are also interested in incorporating users' and their peers' immunity from prior infections. Currently, our calculations do not account for the impact of waning immunity from vaccines or natural infections or how variants of concern may affect the rate of decline. We plan to include these factors in our next release. Moreover, we have not factored in the risk of myocarditis associated with vaccination or infection, a concern that many people have asked us to address. Incorporating this information may demonstrate that the risk of myocarditis is higher with SARS-CoV-2 infection than with vaccination.

While there are areas for improvement, our risk calculator has a wide range of features. Through studies conducted with the risk calculation system, we have found that there are significant synergies between the use of masks with different fitted filtration efficacy (FFE) and different dosages of vaccination, resulting in a decrease in risk. Our risk calculator takes into account all significant variants currently known, including alpha, beta, gamma, delta, and omicron, as well as the increase in risks associated with the presence of a particular variant in a particular region/country. The system updates regularly to fetch the latest number of confirmed cases and variants for 203 countries, making it a reusable tool for the future. The risk calculator has a minimalistic design and can be translated into a web application, making it highly accessible and usable even with slow internet connections. These features make the system highly scalable for users worldwide.

\textbf{Conclusion:} The CovARC COVID-19 risk calculator is a valuable and accurate tool that can assist the public and policymakers in assessing the range of risks of infection, hospitalization, and death from COVID-19. The calculator is more comprehensive and simpler than existing alternatives, with a streamlined interface that allows users to obtain risk estimates quickly. The calculator takes into account all significant factors, including age, gender, health conditions, vaccination status, use of face masks, and the number of people in close contact. Although there is always uncertainty in risk estimation, the calculator provides a range of risk estimates rather than an absolute value. We further plan to improve the calculator by including factors such as previous infection, immunity, waning immunity, and the risk of myocarditis with vaccination or infection. The calculator is regularly updated to include the latest information on variants and confirmed cases for over 203 countries, making it a reusable and highly scalable tool. We hope that this calculator will empower the public to live their lives safely and that a beneficial added effect of using the calculator will be education about risk reduction, which ultimately could result in reduced COVID-19 cases. Future directions will include quantifying the impact of CovARC on COVID-19 community transmission levels. Gamifying risk reduction could be a valuable public health strategy for this and future Pandemics. 

\section*{Methods}
\textbf{Dataset usage:} The risk calculation system uses a variety of different datasets that are extracted from numerous online sources. Our tool carries out a 14-day aggregate in order to estimate the number of active cases. Therefore, the confirmed cases dataset is updated on a daily basis. As for the variants dataset, we take the value from 31 days ago due to the prevalence of a variant for 1 month in the population. We initially carry out pre-processing of the number of confirmed cases obtained through the Johns Hopkins dataset \cite{CSSEGISandData}. We then identify the number of active cases by subtracting the confirmed cases for a given day from the confirmed cases for the previous day and compute a 14-day aggregate for the number of active cases. We carry out a 14-day aggregate due to the fact that the SARS-CoV-2 virus is active for 14 days in any specific condition\cite{Khadem_Charvadeh2022-oy}. We determine a range of 14-day aggregate active cases by taking the upper limit as the product of the 14-day aggregate active cases and the ratio between the confirmed cases obtained from Facebook's COVID-19 Trends and Impact Survey\cite{facebooksurveys} and Johns Hopkins dataset \cite{CSSEGISandData}. Following this, we extract the variants data using the CDC variants dataset \cite{cdcvariants} for calculating the number of different variant cases present in a particular region/country. We estimate the prevalence of a particular variant by carrying out Gaussian smoothing\cite{LIN19961247} taking the variant values 30 days before the current date to account for uneven and delayed data reporting\cite{Florensa2022}.
\begin{table}[ht]
\scalebox{0.55}{
\begin{tabular}{|l|l|}

\hline
\textbf{A.} | {\textbf{Mask Type}}                     & \textbf{FFE} \\ \hline
2-layer woven nylon mask without nose bridge                 & 0.447        \\ \hline
2-layer woven nylon mask with nose bridge                    & 0.563        \\ \hline
2-layer woven nylon  with nose bridge and filter insert      & 0.744        \\ \hline
2-layer woven nylon  with nose bridge washed                 & 0.79         \\ \hline
Cotton Bandana folded surgeon general style                  & 0.49         \\ \hline
Cotton Bandana folded bandit style                           & 0.49         \\ \hline
Single-layer woven polyester gaiter                          & 0.378        \\ \hline
Single-layer woven polyester mask with ties                  & 0.393        \\ \hline
Non-woven polypropylene mask with fixed ear loops            & 0.286        \\ \hline
3-layer knitted cotton mask with ear loops                   & 0.265        \\ \hline
N95 respirator                                               & 0.984        \\ \hline
Surgical mask with ties                                      & 0.715        \\ \hline
Procedure mask with ear loops                                & 0.385        \\ \hline
Procedure mask with loops tied, corners tucked               & 0.603        \\ \hline
Procedure mask with loops tied, corners tucked and ear guard & 0.617        \\ \hline
Procedure mask with Clawed hair clip                         & 0.648        \\ \hline
Procedure mask with fix-the-mask technique (rubber bands)    & 0.782        \\ \hline
Procedure mask with Nylon hosiery sleeve                     & 0.802        \\ \hline
No Mask                                                      & 0            \\ \hline
\end{tabular}}
\scalebox{0.7}{
\begin{tabular}{|l|l|l|l|l|l|l|}
\hline
\textbf{B.} | \textbf{Vaccine} & \textbf{Normal} & \textbf{Alpha} & \textbf{Beta} & \textbf{Gamma} & \textbf{Delta} & \textbf{Omicron} \\ 
\hline
Pfizer (Dose 1) & 0.8 - 0.91 & 0.49 & 0.36-0.375 & 0.36-0.37 & 0.33 & - \\ 
\hline
Pfizer (Dose 2) & 0.95 & 0.87-0.95 & 0.72-0.85 & 0.75-0.77 & 0.79 - 0.92 & 0.07-0.1 \\
\hline
Pfizer (Dose 2) + Pfizer Booster & 0.95 & 0.87-0.95 & 0.72-0.85 & 0.75-0.77 & 0.79-0.92 & 0.44-0.47 \\ \hline
Pfizer (Dose 2) + Moderna (Booster) & 0.95 & 0.87-0.95 & 0.72-0.85 & 0.75-0.77 & 0.79-0.92 & 0.63-0.66 \\ \hline
Moderna (Dose 1) & 0.8-0.9 & 0.49 & 0.72 & 0.72 & 0.33 & - \\ \hline
Moderna (Dose 2) & 0.9-0.96 & 0.91-0.96 & 0.9-0.96 & 0.9-0.96 & 0.855-0.96 & 0.35-0.52 \\ \hline
J\&J (Dose 1) & 0.69-0.77 & 0.77 & 0.52-0.57 & 0.51-0.68 & 0.49-0.78 & - \\ \hline
J\&J (Dose 1) + J\&J Booster & 0.69-0.77 & 0.77 & 0.52-0.57 & 0.51-0.68 & 0.49-0.78 & 0.85 \\ \hline
Astrazeneca (Dose 1) & 0.55-0.67 & 0.33-0.37 & 0.1-0.11 & 0.11-0.243 & 0.329 & - \\ \hline
Astrazeneca (Dose 2) & 0.82-0.85 & 0.66-0.74 & 0.22-0.49 & 0.22-0.49 & 0.59 & - \\ \hline
Astrazeneca (Dose 2) + Pfizer/Moderna (Booster) & 0.82-0.85 & 0.66-0.74 & 0.22-0.7 & 0.22-0.49 & 0.59 & 0.59-0.62\\ \hline
Novavax (Dose 1) & 0.904 & 0.863 & 0.486 & - & - & - \\ \hline
No Vaccine & 0 & 0 & 0 & 0 & 0 & 0\\ \hline
\end{tabular}}
\caption{\small{\label{tab:mask_ffe_eff} A. The first table represents fitted filtration efficacy (FFE)\cite{10.1001/jamainternmed.2020.4221, 10.1001/jamainternmed.2020.8168, maskffearticle} for different types of masks currently available on the market. These values are used in order to calculate the reduction in different risks associated with COVID-19 when different types of masks are used. B. The second table represents the efficacy\cite{Hansen2021.12.20.21267966, Andrews2021.12.14.21267615, Ulloa2021.12.24.21268382, NAABER2021100208, Goel2021.08.23.457229, doi:10.1126/science.abj4176, Khoury2021, Rella2021, doi:10.1056/NEJMoa2110345, yourlocalepimay2021} of different types of vaccines against COVID-19 and its variants. This table is used in order to identify reduction is risks associated with COVID-19 when different types of vaccines and dosage is taken.}}
\end{table}

\begin{table}[ht]
\scalebox{0.7}{
\begin{tabular}{|l|l|l|}
\hline
 & \textbf{Hospitalization Rate (\%)} & \textbf{Death Rate (\%)} \\
\hline
\textbf{By Age Group} & & \\
\hline
0-17 years old & 0.8\% & 0.0015\% \\
\hline
18-49 years old & 2.5\% & 0.07\% \\
\hline
50-64 years old & 7.9\% & 0.7\% \\
\hline
65+ years old & 23\% & 6\% \\
\hline
All ages & 5\% & 0.75\% \\
\hline
\textbf{By SARS-CoV-2 Variant- Fold Higher Risk Compared to Original Variants} & & \\
\hline
Alpha (B.1.1.7, B.1.1.7 with E484K\cite{Yang2022-na}) & 1.5(1.5-1.6) & 1.6(1.4-1.7) \\
\hline
Beta (B.1.351) & Under Investigation & Possibly Increased \\
\hline
Gamma (P.1) & Possibly Increased & 1.5 (1.2-1.9) \\
\hline
Delta (B.1.617.2) & 2.3 (1.9-3.0) & 2.4 (1.5-3.3) \\
\hline
\textbf{By Gender - Fold Higher Risk} & & \\
\hline
Male & - & 1.5-2.3 \\
\hline
Female & - & 1 \\
\hline
\textbf{By any chronic health condition - Fold Higher Risk} & 2.5 & 1.2-6.9 \\
\hline
\end{tabular}}
\caption{\small{\label{tab:hosp_death}The given table explains the influence of Age group\cite{Cannistraci2021}, presence of a specific variant\cite{Zali2022}, gender of the user as well as past chronic illness/health conditions (Diabetes, Heart Disease, Cancer, Lung disease, High Blood Pressure, Immunocompromised, Asthma, Kidney Disease, Obesity, Sickle Cell Anemia, HIV, Liver Disease) on risk of hospitalization and death\cite{Aslaner2021-yx}. These factors are taken into account in our risk calculator when we calculate the risk of hospitalization and risk of death from the risk of infection for an individual\cite{10.1093/infdis/jiac063}.}}
\end{table}
In addition to the variants and confirmed cases datasets that we obtain from online sources, we create custom datasets where we used information provided by several different research studies and articles. These datasets consiss o f the mask's fitted filtration efficacy (FFE) dataset (see \href{tab:mask_ffe_eff}{Table \ref{tab:mask_ffe_eff} A.}) and the efficacy of the vaccine against different variants of virus dataset (see \href{tab:mask_ffe_eff}{Table \ref{tab:mask_ffe_eff} B.}) which we use in order to estimate risk reduction. The vaccine efficacy dataset is coupled with different variants that are extracted using the GISAID dataset to estimate the overall impact of variants with a specific level of vaccination or no vaccination. The mask dataset is used to estimate the overall reduction in risk.

In order to add the influence of indoor and outdoor environment, age, gender, past chronic illness, and variants on risk of infection, hospitalization and death, we further add a multiplier which helps us to accurate the risks associated with carrying out a specific activity during the COVID-19 pandemic. We use all of these datasets and consider the number of people the user passes outdoors while traveling to the location and the number of people they are with at the destination (which can be indoors or outdoors) to estimate several risks associated with COVID-19 (see \href{tab:hosp_death}{Table \ref{tab:hosp_death}}). 

\textbf{Pipeline and Workflow:}

\begin{figure}[ht]
\includegraphics[width=1\textwidth]{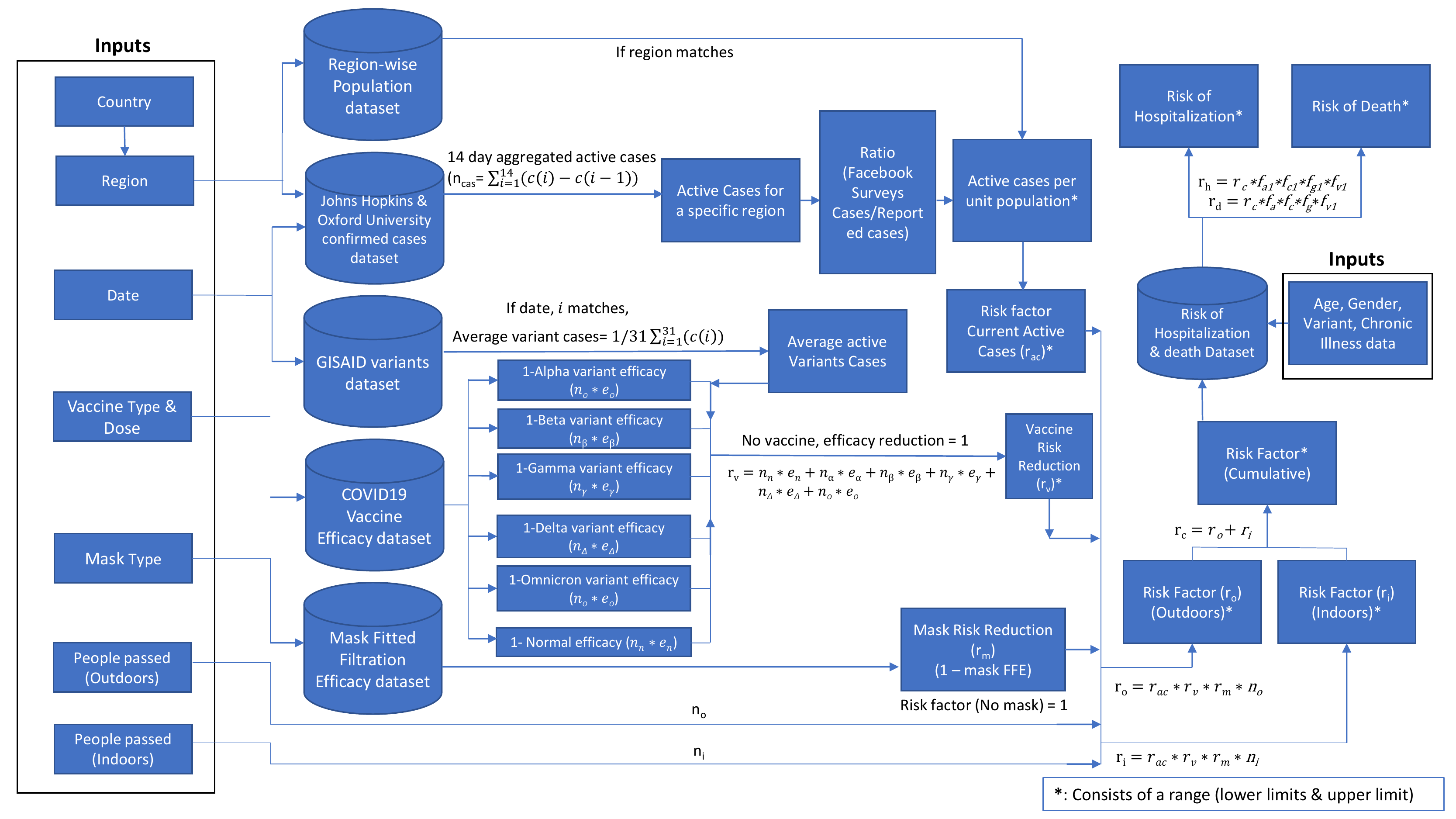}
\caption{\small{This figure represents a simplified workflow diagram of the risk calculation along with different datasets as well as other inputs used in order to estimate different risks related to COVID-19. The diagram also represents the steps that are performed when calculating the risk reduction when different preventive measures are taken}}
\label{fig:wireframe}
\end{figure}

As shown in \href{fig:wireframe}{Figure \ref{fig:wireframe}}, we then identify the new cases from the confirmed cases by subtracting the confirmed cases for a given day from the confirmed cases for the previous day. We then identify the number of new cases by taking a 14-day sum for the number of active cases($n_{ac}$). Upon providing the input for region/state within the country and present-day date, we identify the 14-day sum of the active case for a particular previous day and region/state. We then consider the region and date to identify the variants and the average number of variant cases over 30 days. We use the ratio between the confirmed cases reported through Facebook surveys\cite{facebooksurveys} and the officially reported confirmed cases and multiply them with the confirmed cases reported by Johns Hopkins University dataset\cite{CSSEGISandData} in order to estimate the upper limit for the number of reported cases.  

We calculated the reduced risk of infection for variants of SARS-CoV-2 by subtracting one from the efficacy of vaccines and fitted filtration efficacy of masks using the same method. For vaccines, we use the lower and upper limits of efficacy to estimate and multiply it with the higher and lower limits of confirmed cases, respectively and estimate the risk of infection. We then calculate the risk of variants by considering the regional data to identify the number of variants present and multiply them with reduced risk after vaccination to identify the added risk of variants ($r_{v}$). Furthermore, we multiply the number of people passed by indoors ($n_{i}$), outdoors ($n_{o}$), risk due to active cases per unit population ($r_{ac}$), risk reduction due to vaccination ($r_{v}$) and reduced risk due to mask ($r_{m}$) to calculate indoor ($r_{i}$) and outdoor risk ($r_{o}$) range of infection. The sum of this helps us to estimate the range of cumulative risk of infection.

We further estimate the risk of hospitalization ($r_{h}$) and risk of death ($r_{d}$) by using factors related to age ($f_{a}$ and $f_{a1}$), gender($f_{g}$ and $f_{g1}$), past chronic illness($f_{c}$ and $f_{c1}$) and type of variants ($f_{v}$ and $f_{v1}$) and multiply the range of risk factors with the upper and lower limits of the cumulative risk of infection. Finally, we use it to compute the range of risk of hospitalization and risk of death.
\begin{figure}[ht]
\centering
\includegraphics[width=1\textwidth]{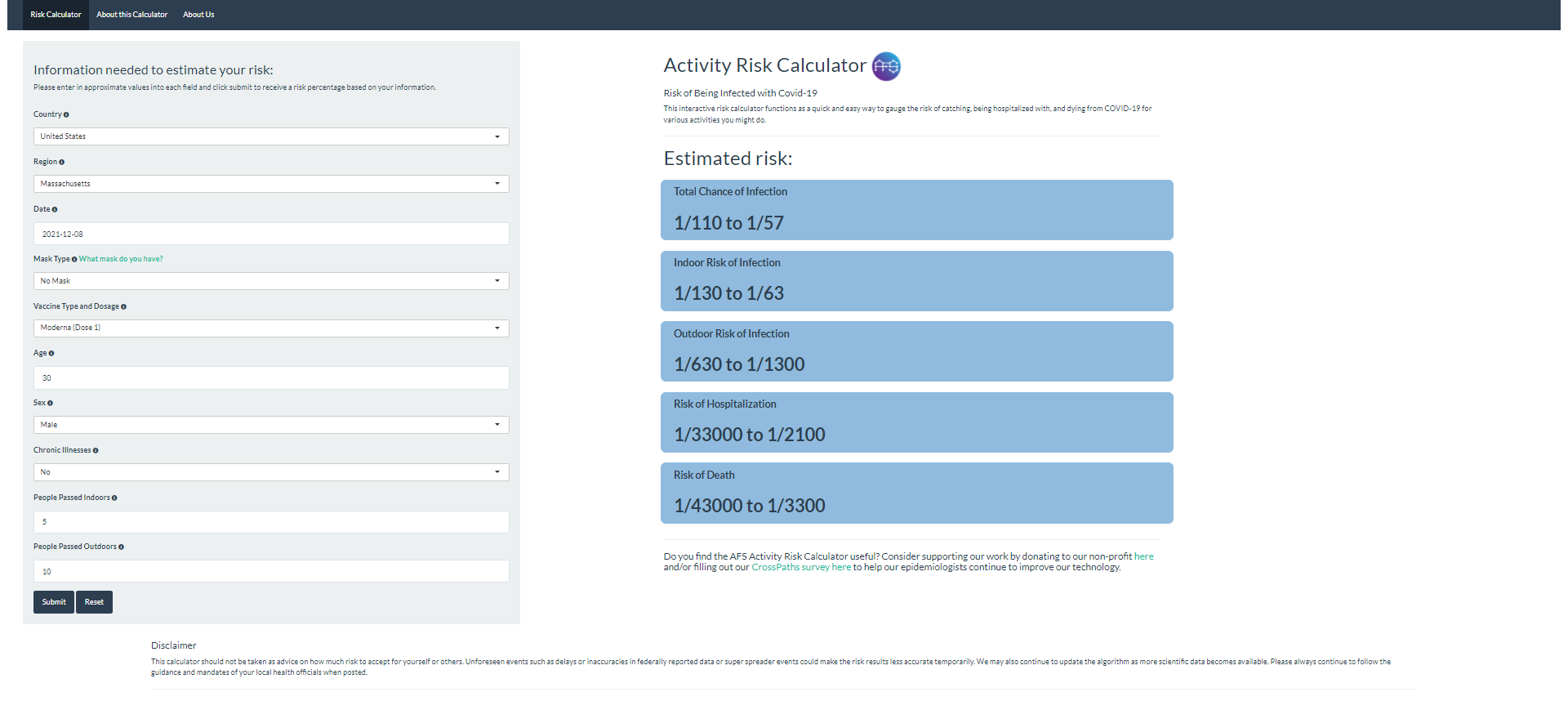}
\caption{\small{The given image is a screenshot of the user interface of the web application that was translated using the risk calculation system. The web application uses a form-like structure with minimal fields to ensure that users can carry out risk estimation as quickly as possible. After clicking the submit button, the results are displayed on the right partition of the web application. Using this, the user can identify different personal risks for carrying out a specific daily life activity}}
\label{fig:webpage}
\end{figure}

\textbf{Testing \& Deployment of Web Application:}
Upon finishing the implementation of the risk calculation system as well as several studies related to its robustness, we proceeded to create a web application that could enable the risk calculation system to be user-facing. We used shinyapps\cite{rshiny} to translate the R code into a web application by creating a simple form-like user interface. Using Rshiny also enabled us to ensure that the webpage is lightweight and can be used with a slow internet connection as well. With this tool, we provided a channel where anyone could access the risk estimation system to identify the personal risk of infection, hospitalization, and death when carrying out daily activities. In order to make it more accessible, the web page's user interface was also translated for smartphones. This is useful to increase the accessibility of the risk calculator web application to the general public and enable smooth implementation of the system in the daily life of individuals. The representation of the webpage can be seen \href{fig:webpage}{Figure \ref{fig:webpage}}.

\section*{Author contributions statement}
Shreyasvi Natraj created the code for the backend of the calculator, made the paper figures, and wrote a large part of the paper draft. Nathan Yap interpreted Shreyasvi's code and created the user interface in R Shiny and deployed the web application that could be used by the general public. Malhar Bhide researched the effects of interventions, vaccines, and masks as well as drew the comparison between our risk calculator and presently existing COVID-19 risk calculators. He further contributed to creating a few tables in the research paper. Meng Liu created the variants data-extraction pipeline and performed Gaussian smoothing. Christin Glorioso was responsible for providing the idea of the risk calculation and contributed to reviewing the manuscript. She also was responsible for reviewing the code for the risk calculator as well as providing feedback to improve the web application along with Agrima Seth and Jonathan Berman. 

\section*{Availibility of Data \& Materials}
All of the data used in this study is publicly available and can be either directly accessed or accessed upon filing a request. The data related to the number of active, confirmed, and death cases due to COVID-19 is maintained by the Center for Systems Science and Engineering (CSSE) at Johns Hopkins University at \href{COVID-19 Data Repository by the Center for Systems Science and Engineering (CSSE) at Johns Hopkins University} {https://github.com/CSSEGISandData/COVID-19}. The data for the SARS-CoV-2 variants can be obtained upon request from the repository maintained by  Global Initiative on Sharing Avian Influenza Data (GISAID) Initiative \href{Global Initiative on Sharing Avian Influenza Data}{https://gisaid.org/hcov19-variants}. Facebook/Meta COVID-19 Data used for cross-validation of the number of confirmed cases can be accessed at \href{Data for Good}{https://dataforgood.facebook.com/dfg/tools/covid-19-trends-and-impact-survey}. Factors used for calculating risk reduction by use of masks and different dosages of vaccines can be seen in \href{tab:mask_ffe_eff}{Table \ref{tab:mask_ffe_eff} A.} and \href{tab:mask_ffe_eff}{Table \ref{tab:mask_ffe_eff} B.}. Factors used for calculating the risk of hospitalization and death can be seen in \href{tab:hosp_death}{Table \ref{tab:hosp_death}}. The source code that supports the findings of this research is available from the corresponding author upon request.
 
\bibliography{sample}

\begin{thebibliography}{10}
\urlstyle{rm}
\expandafter\ifx\csname url\endcsname\relax
  \def\url#1{\texttt{#1}}\fi
\expandafter\ifx\csname urlprefix\endcsname\relax\def\urlprefix{URL }\fi
\expandafter\ifx\csname doiprefix\endcsname\relax\def\doiprefix{DOI: }\fi
\providecommand{\bibinfo}[2]{#2}
\providecommand{\eprint}[2][]{\url{#2}}

\bibitem{Martinez-Perez2020-va}
\bibinfo{author}{Martinez-Perez, C.}, \bibinfo{author}{Alvarez-Peregrina, C.},
  \bibinfo{author}{Villa-Collar, C.} \& \bibinfo{author}{S{\'a}nchez-Tena,
  M.~{\'A}.}
\newblock \bibinfo{journal}{\bibinfo{title}{Citation network analysis of the
  novel coronavirus disease 2019 ({COVID-19})}}.
\newblock {\emph{\JournalTitle{Int. J. Environ. Res. Public Health}}}
  \textbf{\bibinfo{volume}{17}}, \bibinfo{pages}{7690} (\bibinfo{year}{2020}).

\bibitem{Subramanian2022}
\bibinfo{author}{Subramanian, A.} \emph{et~al.}
\newblock \bibinfo{journal}{\bibinfo{title}{Symptoms and risk factors for long
  covid in non-hospitalized adults}}.
\newblock {\emph{\JournalTitle{Nature Medicine}}}
  \textbf{\bibinfo{volume}{28}}, \bibinfo{pages}{1706--1714},
  \doiprefix\url{10.1038/s41591-022-01909-w} (\bibinfo{year}{2022}).

\bibitem{roy_ghosh}
\bibinfo{author}{Roy, S.} \& \bibinfo{author}{Ghosh, P.}
\newblock \bibinfo{title}{Factors affecting covid-19 infected and death rates
  inform lockdown-related policymaking}.

\bibitem{10.1371/journal.pmed.1003871}
\bibinfo{author}{Bhaskaran, K.} \emph{et~al.}
\newblock \bibinfo{journal}{\bibinfo{title}{Overall and cause-specific
  hospitalisation and death after covid-19 hospitalisation in england: A cohort
  study using linked primary care, secondary care, and death registration data
  in the opensafely platform}}.
\newblock {\emph{\JournalTitle{PLOS Medicine}}} \textbf{\bibinfo{volume}{19}},
  \bibinfo{pages}{1--20}, \doiprefix\url{10.1371/journal.pmed.1003871}
  (\bibinfo{year}{2022}).

\bibitem{doi:10.1177/1178633721991260}
\bibinfo{author}{Sanyaolu, A.} \emph{et~al.}
\newblock \bibinfo{journal}{\bibinfo{title}{Global pandemicity of covid-19:
  Situation report as of june 9, 2020}}.
\newblock {\emph{\JournalTitle{Infectious Diseases: Research and Treatment}}}
  \textbf{\bibinfo{volume}{14}}, \bibinfo{pages}{1178633721991260},
  \doiprefix\url{10.1177/1178633721991260} (\bibinfo{year}{2021}).
\newblock \bibinfo{note}{PMID: 33597811},
  \eprint{https://doi.org/10.1177/1178633721991260}.

\bibitem{E_Vincenzo2021-vu}
\bibinfo{author}{{Vincenzo}} \emph{et~al.}
\newblock \bibinfo{journal}{\bibinfo{title}{Indoor versus outdoor transmission
  of {SARS-COV-2}: environmental factors in virus spread and underestimated
  sources of risk}}.
\newblock {\emph{\JournalTitle{EuroMediterr J Environ Integr}}}
  \textbf{\bibinfo{volume}{6}}, \bibinfo{pages}{30} (\bibinfo{year}{2021}).

\bibitem{Haas2020-fd}
\bibinfo{author}{Haas, C.}
\newblock \bibinfo{journal}{\bibinfo{title}{Coronavirus and risk analysis}}.
\newblock {\emph{\JournalTitle{Risk Anal.}}} \textbf{\bibinfo{volume}{40}},
  \bibinfo{pages}{660--661} (\bibinfo{year}{2020}).

\bibitem{microcovidproject}
\bibinfo{title}{The microcovid project}.
\newblock \bibinfo{howpublished}{\url{https://www.microcovid.org/paper/all}}.
\newblock \bibinfo{note}{Accessed: 2021-10-29}.

\bibitem{nytimesarticle}
\bibinfo{title}{California covid restrictions}.
\newblock
  \bibinfo{howpublished}{\url{https://www.nytimes.com/2020/12/04/briefing/california-covid-restrictions-warner-bros-stimulus.html
  }} (\bibinfo{year}{2021}).

\bibitem{infodemic}
\bibinfo{author}{Hu, Z.}, \bibinfo{author}{Yang, Z.}, \bibinfo{author}{Li, Q.}
  \& \bibinfo{author}{Zhang, A.}
\newblock \bibinfo{journal}{\bibinfo{title}{The covid-19 infodemic:
  Infodemiology study analyzing stigmatizing search terms}}.
\newblock {\emph{\JournalTitle{J Med Internet Res}}}
  \textbf{\bibinfo{volume}{22}}, \bibinfo{pages}{e22639},
  \doiprefix\url{10.2196/22639} (\bibinfo{year}{2020}).

\bibitem{evaluation_2022}
\bibinfo{author}{Luu, M.~N.} \emph{et~al.}
\newblock \bibinfo{journal}{\bibinfo{title}{Evaluation of risk factors
  associated with sars-cov-2 transmission}}.
\newblock {\emph{\JournalTitle{Current Medical Research and Opinion}}}
  \textbf{\bibinfo{volume}{38}}, \bibinfo{pages}{2021--2028},
  \doiprefix\url{10.1080/03007995.2022.2125258} (\bibinfo{year}{2022}).
\newblock \bibinfo{note}{PMID: 36106710},
  \eprint{https://doi.org/10.1080/03007995.2022.2125258}.

\bibitem{coviddatatracker}
\bibinfo{author}{for Disease~Control, C.} \& \bibinfo{author}{Prevention}.
\newblock \bibinfo{title}{Covid data tracker}.
\newblock
  \bibinfo{howpublished}{\url{https://covid.cdc.gov/covid-data-tracker/\#county-view&list_select_map_data_parent=Risk&map-metrics-cv-comm-transmission=community_transmission_level&list_select_map_data_metro=all}}
  (\bibinfo{year}{2019}).

\bibitem{qcovid}
\bibinfo{title}{Qcovid risk calculator}.
\newblock
  \bibinfo{howpublished}{\url{https://qcovid.org/Home/AcademicLicence?licencedUrl=\%2FCalculation}}.
\newblock \bibinfo{note}{Accessed: 2021-10-29}.

\bibitem{ASIMI}
\bibinfo{title}{Covid-19 risk calculator}.
\newblock
  \bibinfo{howpublished}{\url{https://covid-19.forhealth.org/covid-19-transmission-calculator/}}.
\newblock \bibinfo{note}{Accessed: 2021-10-29}.

\bibitem{19andme}
\bibinfo{title}{19 and me: Covid-19 risk score calculator}.
\newblock
  \bibinfo{howpublished}{\url{https://19andme.covid19.mathematica.org/}}.
\newblock \bibinfo{note}{Accessed: 2021-10-29}.

\bibitem{mycovidrisk}
\bibinfo{title}{Mycovidrisk}.
\newblock \bibinfo{howpublished}{\url{https://mycovidrisk.app/}}.
\newblock \bibinfo{note}{Accessed: 2021-10-29}.

\bibitem{covidgatech}
\bibinfo{title}{Covid-19 event risk assessment planning tool}.
\newblock \bibinfo{howpublished}{\url{https://covid19risk.biosci.gatech.edu/}}.
\newblock \bibinfo{note}{Accessed: 2021-10-29}.

\bibitem{covidtracker}
\bibinfo{title}{Covid19 exposure risk calculator}.
\newblock
  \bibinfo{howpublished}{\url{https://covidtracker.fr/covid19-risk-calculator/}}.
\newblock \bibinfo{note}{Accessed: 2021-10-29}.

\bibitem{maxplancktracker}
\bibinfo{author}{Lelieveld, J.} \emph{et~al.}
\newblock \bibinfo{journal}{\bibinfo{title}{Model calculations of aerosol
  transmission and infection risk of covid-19 in indoor environments}}.
\newblock {\emph{\JournalTitle{International Journal of Environmental Research
  and Public Health}}} \textbf{\bibinfo{volume}{17}},
  \doiprefix\url{10.3390/ijerph17218114} (\bibinfo{year}{2020}).

\bibitem{mitriskcalculator}
\bibinfo{author}{Bazant, M.~Z.} \& \bibinfo{author}{Bush, J. W.~M.}
\newblock \bibinfo{journal}{\bibinfo{title}{A guideline to limit indoor
  airborne transmission of covid-19}}.
\newblock {\emph{\JournalTitle{Proceedings of the National Academy of
  Sciences}}} \textbf{\bibinfo{volume}{118}},
  \doiprefix\url{10.1073/pnas.2018995118} (\bibinfo{year}{2021}).
\newblock \eprint{https://www.pnas.org/content/118/17/e2018995118.full.pdf}.

\bibitem{northwestedu}
\bibinfo{title}{Covid-19 dashboard}.
\newblock
  \bibinfo{howpublished}{\url{https://www.northwestern.edu/coronavirus-covid-19-updates/university-status/dashboard/
  }} (\bibinfo{year}{2021}).

\bibitem{jin2020assessment}
\bibinfo{author}{Jin, J.} \emph{et~al.}
\newblock \bibinfo{journal}{\bibinfo{title}{Assessment of individual-and
  community-level risks for covid-19 mortality in the us and implications for
  vaccine distribution}}.
\newblock {\emph{\JournalTitle{medRxiv}}}  (\bibinfo{year}{2020}).

\bibitem{Rowe2020.12.30.20249058}
\bibinfo{author}{Rowe, B.}, \bibinfo{author}{Canosa, A.},
  \bibinfo{author}{Drouffe, J.} \& \bibinfo{author}{Mitchell, J.}
\newblock \bibinfo{journal}{\bibinfo{title}{Simple quantitative assessment of
  the outdoor versus indoor airborne transmission of viruses and covid-19}}.
\newblock {\emph{\JournalTitle{medRxiv}}}
  \doiprefix\url{10.1101/2020.12.30.20249058} (\bibinfo{year}{2021}).
\newblock
  \eprint{https://www.medrxiv.org/content/early/2021/01/04/2020.12.30.20249058.1.full.pdf}.

\bibitem{CSSEGISandData}
\bibinfo{author}{Gardner, L.}
\newblock \bibinfo{title}{Covid-19 data repository}.
\newblock
  \bibinfo{howpublished}{\url{https://github.com/CSSEGISandData/COVID-19}}
  (\bibinfo{year}{2019}).

\bibitem{Zelenova2021-lt}
\bibinfo{author}{Zelenova, M.}, \bibinfo{author}{Ivanova, A.},
  \bibinfo{author}{Semyonov, S.} \& \bibinfo{author}{Gankin, Y.}
\newblock \bibinfo{journal}{\bibinfo{title}{Analysis of 329,942 {SARS-CoV-2}
  records retrieved from {GISAID} database}}.
\newblock {\emph{\JournalTitle{Comput. Biol. Med.}}}
  \textbf{\bibinfo{volume}{139}}, \bibinfo{pages}{104981}
  (\bibinfo{year}{2021}).

\bibitem{surveys}
\bibinfo{author}{Astley, C.~M.} \emph{et~al.}
\newblock \bibinfo{journal}{\bibinfo{title}{Global monitoring of the impact of
  the covid-19 pandemic through online surveys sampled from the facebook user
  base}}.
\newblock {\emph{\JournalTitle{Proceedings of the National Academy of
  Sciences}}} \textbf{\bibinfo{volume}{118}}, \bibinfo{pages}{e2111455118},
  \doiprefix\url{10.1073/pnas.2111455118} (\bibinfo{year}{2021}).
\newblock \eprint{https://www.pnas.org/doi/pdf/10.1073/pnas.2111455118}.

\bibitem{cdcvariants}
\bibinfo{author}{for Disease~Control, C.}, \bibinfo{author}{Prevention}
  \emph{et~al.}
\newblock \bibinfo{journal}{\bibinfo{title}{Variant proportions}}.
\newblock {\emph{\JournalTitle{Retrieved November}}}
  \textbf{\bibinfo{volume}{16}}, \bibinfo{pages}{2020} (\bibinfo{year}{2021}).

\bibitem{Wu2020-ju}
\bibinfo{author}{Wu, Y.-C.}, \bibinfo{author}{Chen, C.-S.} \&
  \bibinfo{author}{Chan, Y.-J.}
\newblock \bibinfo{journal}{\bibinfo{title}{The outbreak of {COVID-19}: An
  overview}}.
\newblock {\emph{\JournalTitle{J. Chin. Med. Assoc.}}}
  \textbf{\bibinfo{volume}{83}}, \bibinfo{pages}{217--220}
  (\bibinfo{year}{2020}).

\bibitem{maskffearticle}
\bibinfo{author}{Agency, U. E.~P.}
\newblock \bibinfo{title}{Epa researchers test effectiveness of face masks,
  disinfection methods against covid-19}.
\newblock
  \bibinfo{howpublished}{\url{https://www.epa.gov/sciencematters/epa-researchers-test-effectiveness-face-masks-disinfection-methods-against-covid-19}}
  (\bibinfo{year}{2021}).

\bibitem{useofmaskscdcarticle}
\bibinfo{author}{for Disease~Control, C.} \& \bibinfo{author}{Prevention}.
\newblock \bibinfo{title}{Science brief: Community use of masks to control the
  spread of sars-cov-2}.
\newblock
  \bibinfo{howpublished}{\url{https://www.cdc.gov/coronavirus/2019-ncov/science/science-briefs/masking-science-sars-cov2.html}}
  (\bibinfo{year}{2021}).

\bibitem{riskcalculatorlink}
\bibinfo{title}{Covid19 activity risk calculator webpage}.
\newblock
  \bibinfo{howpublished}{\url{https://realsciencecommunity.shinyapps.io/riskcalculator/}}
  (\bibinfo{year}{2021}).

\bibitem{rshiny}
\bibinfo{author}{Chang, W.} \emph{et~al.}
\newblock \emph{\bibinfo{title}{shiny: Web Application Framework for R}}
  (\bibinfo{year}{2022}).
\newblock \bibinfo{note}{R package version 1.7.3.9001}.

\bibitem{Khadem_Charvadeh2022-oy}
\bibinfo{author}{Khadem~Charvadeh, Y.}, \bibinfo{author}{Yi, G.~Y.},
  \bibinfo{author}{Bian, Y.} \& \bibinfo{author}{He, W.}
\newblock \bibinfo{journal}{\bibinfo{title}{Is 14-days a sensible quarantine
  length for {COVID-19}? examinations of some associated issues with a case
  study of {COVID-19} incubation times}}.
\newblock {\emph{\JournalTitle{Stat. Biosci.}}} \textbf{\bibinfo{volume}{14}},
  \bibinfo{pages}{175--190} (\bibinfo{year}{2022}).

\bibitem{facebooksurveys}
\bibinfo{title}{Covid-19 trends and impact survey}.
\newblock
  \bibinfo{howpublished}{\url{https://dataforgood.facebook.com/dfg/tools/covid-19-trends-and-impact-survey}}
  (\bibinfo{year}{2021}).

\bibitem{LIN19961247}
\bibinfo{author}{Lin, H.-C.}, \bibinfo{author}{Wang, L.-L.} \&
  \bibinfo{author}{Yang, S.-N.}
\newblock \bibinfo{journal}{\bibinfo{title}{Automatic determination of the
  spread parameter in gaussian smoothing}}.
\newblock {\emph{\JournalTitle{Pattern Recognition Letters}}}
  \textbf{\bibinfo{volume}{17}}, \bibinfo{pages}{1247--1252},
  \doiprefix\url{https://doi.org/10.1016/0167-8655(96)00082-7}
  (\bibinfo{year}{1996}).

\bibitem{Florensa2022}
\bibinfo{author}{Florensa, D.} \emph{et~al.}
\newblock \bibinfo{journal}{\bibinfo{title}{Severity of covid-19 cases in the
  months of predominance of the alpha and delta variants}}.
\newblock {\emph{\JournalTitle{Scientific Reports}}}
  \textbf{\bibinfo{volume}{12}}, \bibinfo{pages}{15456},
  \doiprefix\url{10.1038/s41598-022-19125-4} (\bibinfo{year}{2022}).

\bibitem{10.1001/jamainternmed.2020.4221}
\bibinfo{author}{Sickbert-Bennett, E.~E.} \emph{et~al.}
\newblock \bibinfo{journal}{\bibinfo{title}{{Filtration Efficiency of Hospital
  Face Mask Alternatives Available for Use During the COVID-19 Pandemic}}}.
\newblock {\emph{\JournalTitle{JAMA Internal Medicine}}}
  \textbf{\bibinfo{volume}{180}}, \bibinfo{pages}{1607--1612},
  \doiprefix\url{10.1001/jamainternmed.2020.4221} (\bibinfo{year}{2020}).
\newblock
  \eprint{https://jamanetwork.com/journals/jamainternalmedicine/articlepdf/2769443/jamainternal\_sickbertbennett\_2020\_oi\_200067\_1606936139.25622.pdf}.

\bibitem{10.1001/jamainternmed.2020.8168}
\bibinfo{author}{Clapp, P.~W.} \emph{et~al.}
\newblock \bibinfo{journal}{\bibinfo{title}{{Evaluation of Cloth Masks and
  Modified Procedure Masks as Personal Protective Equipment for the Public
  During the COVID-19 Pandemic}}}.
\newblock {\emph{\JournalTitle{JAMA Internal Medicine}}}
  \textbf{\bibinfo{volume}{181}}, \bibinfo{pages}{463--469},
  \doiprefix\url{10.1001/jamainternmed.2020.8168} (\bibinfo{year}{2021}).
\newblock
  \eprint{https://jamanetwork.com/journals/jamainternalmedicine/articlepdf/2774266/jamainternal\_clapp\_2020\_oi\_200108\_1617393569.436.pdf}.

\bibitem{Hansen2021.12.20.21267966}
\bibinfo{author}{Hansen, C.~H.} \emph{et~al.}
\newblock \bibinfo{journal}{\bibinfo{title}{Vaccine effectiveness against
  sars-cov-2 infection with the omicron or delta variants following a two-dose
  or booster bnt162b2 or mrna-1273 vaccination series: A danish cohort study}}.
\newblock {\emph{\JournalTitle{medRxiv}}}
  \doiprefix\url{10.1101/2021.12.20.21267966} (\bibinfo{year}{2021}).
\newblock
  \eprint{https://www.medrxiv.org/content/early/2021/12/22/2021.12.20.21267966.full.pdf}.

\bibitem{Andrews2021.12.14.21267615}
\bibinfo{author}{Andrews, N.} \emph{et~al.}
\newblock \bibinfo{journal}{\bibinfo{title}{Effectiveness of covid-19 vaccines
  against the omicron (b.1.1.529) variant of concern}}.
\newblock {\emph{\JournalTitle{medRxiv}}}
  \doiprefix\url{10.1101/2021.12.14.21267615} (\bibinfo{year}{2021}).
\newblock
  \eprint{https://www.medrxiv.org/content/early/2021/12/14/2021.12.14.21267615.full.pdf}.

\bibitem{Ulloa2021.12.24.21268382}
\bibinfo{author}{Ulloa, A.~C.}, \bibinfo{author}{Buchan, S.~A.},
  \bibinfo{author}{Daneman, N.} \& \bibinfo{author}{Brown, K.~A.}
\newblock \bibinfo{journal}{\bibinfo{title}{Early estimates of sars-cov-2
  omicron variant severity based on a matched cohort study, ontario, canada}}.
\newblock {\emph{\JournalTitle{medRxiv}}}
  \doiprefix\url{10.1101/2021.12.24.21268382} (\bibinfo{year}{2022}).
\newblock
  \eprint{https://www.medrxiv.org/content/early/2022/01/02/2021.12.24.21268382.full.pdf}.

\bibitem{NAABER2021100208}
\bibinfo{author}{Naaber, P.} \emph{et~al.}
\newblock \bibinfo{journal}{\bibinfo{title}{Dynamics of antibody response to
  bnt162b2 vaccine after six months: a longitudinal prospective study}}.
\newblock {\emph{\JournalTitle{The Lancet Regional Health - Europe}}}
  \textbf{\bibinfo{volume}{10}}, \bibinfo{pages}{100208},
  \doiprefix\url{https://doi.org/10.1016/j.lanepe.2021.100208}
  (\bibinfo{year}{2021}).

\bibitem{Goel2021.08.23.457229}
\bibinfo{author}{Goel, R.~R.} \emph{et~al.}
\newblock \bibinfo{journal}{\bibinfo{title}{mrna vaccination induces durable
  immune memory to sars-cov-2 with continued evolution to variants of
  concern}}.
\newblock {\emph{\JournalTitle{bioRxiv}}}
  \doiprefix\url{10.1101/2021.08.23.457229} (\bibinfo{year}{2021}).
\newblock
  \eprint{https://www.biorxiv.org/content/early/2021/08/23/2021.08.23.457229.full.pdf}.

\bibitem{doi:10.1126/science.abj4176}
\bibinfo{author}{Pegu, A.} \emph{et~al.}
\newblock \bibinfo{journal}{\bibinfo{title}{Durability of mrna-1273
  vaccine\&\#x2013;induced antibodies against sars-cov-2 variants}}.
\newblock {\emph{\JournalTitle{Science}}} \textbf{\bibinfo{volume}{373}},
  \bibinfo{pages}{1372--1377}, \doiprefix\url{10.1126/science.abj4176}
  (\bibinfo{year}{2021}).
\newblock \eprint{https://www.science.org/doi/pdf/10.1126/science.abj4176}.

\bibitem{Khoury2021}
\bibinfo{author}{Khoury, D.~S.} \emph{et~al.}
\newblock \bibinfo{journal}{\bibinfo{title}{Neutralizing antibody levels are
  highly predictive of immune protection from symptomatic sars-cov-2
  infection}}.
\newblock {\emph{\JournalTitle{Nature Medicine}}}
  \textbf{\bibinfo{volume}{27}}, \bibinfo{pages}{1205--1211},
  \doiprefix\url{10.1038/s41591-021-01377-8} (\bibinfo{year}{2021}).

\bibitem{Rella2021}
\bibinfo{author}{Rella, S.~A.}, \bibinfo{author}{Kulikova, Y.~A.},
  \bibinfo{author}{Dermitzakis, E.~T.} \& \bibinfo{author}{Kondrashov, F.~A.}
\newblock \bibinfo{journal}{\bibinfo{title}{Rates of sars-cov-2 transmission
  and vaccination impact the fate of vaccine-resistant strains}}.
\newblock {\emph{\JournalTitle{Scientific Reports}}}
  \textbf{\bibinfo{volume}{11}}, \bibinfo{pages}{15729},
  \doiprefix\url{10.1038/s41598-021-95025-3} (\bibinfo{year}{2021}).

\bibitem{doi:10.1056/NEJMoa2110345}
\bibinfo{author}{Thomas, S.~J.} \emph{et~al.}
\newblock \bibinfo{journal}{\bibinfo{title}{Safety and efficacy of the bnt162b2
  mrna covid-19 vaccine through 6 months}}.
\newblock {\emph{\JournalTitle{New England Journal of Medicine}}}
  \textbf{\bibinfo{volume}{385}}, \bibinfo{pages}{1761--1773},
  \doiprefix\url{10.1056/NEJMoa2110345} (\bibinfo{year}{2021}).
\newblock \bibinfo{note}{PMID: 34525277},
  \eprint{https://doi.org/10.1056/NEJMoa2110345}.

\bibitem{yourlocalepimay2021}
\bibinfo{title}{Vaccine table update: May 5, 2021}.
\newblock
  \bibinfo{howpublished}{\url{https://yourlocalepidemiologist.substack.com/p/vaccine-table-update-may-5-2021
  }} (\bibinfo{year}{2021}).

\bibitem{Yang2022-na}
\bibinfo{author}{Yang, W.-T.} \emph{et~al.}
\newblock \bibinfo{journal}{\bibinfo{title}{{SARS-CoV-2} {E484K} mutation
  narrative review: Epidemiology, immune escape, clinical implications, and
  future considerations}}.
\newblock {\emph{\JournalTitle{Infect. Drug Resist.}}}
  \textbf{\bibinfo{volume}{15}}, \bibinfo{pages}{373--385}
  (\bibinfo{year}{2022}).

\bibitem{Cannistraci2021}
\bibinfo{author}{Cannistraci, C.~V.}, \bibinfo{author}{Valsecchi, M.~G.} \&
  \bibinfo{author}{Capua, I.}
\newblock \bibinfo{journal}{\bibinfo{title}{Age-sex population adjusted
  analysis of disease severity in epidemics as a tool to devise public health
  policies for covid-19}}.
\newblock {\emph{\JournalTitle{Scientific Reports}}}
  \textbf{\bibinfo{volume}{11}}, \bibinfo{pages}{11787},
  \doiprefix\url{10.1038/s41598-021-89615-4} (\bibinfo{year}{2021}).

\bibitem{Zali2022}
\bibinfo{author}{Zali, A.} \emph{et~al.}
\newblock \bibinfo{journal}{\bibinfo{title}{Mortality among hospitalized
  covid-19 patients during surges of sars-cov-2 alpha (b.1.1.7) and delta
  (b.1.617.2) variants}}.
\newblock {\emph{\JournalTitle{Scientific Reports}}}
  \textbf{\bibinfo{volume}{12}}, \bibinfo{pages}{18918},
  \doiprefix\url{10.1038/s41598-022-23312-8} (\bibinfo{year}{2022}).

\bibitem{Aslaner2021-yx}
\bibinfo{author}{Aslaner, H.}, \bibinfo{author}{Aslaner, H.~A.},
  \bibinfo{author}{G{\"o}k{\c c}ek, M.~B.}, \bibinfo{author}{Benli, A.~R.} \&
  \bibinfo{author}{Y{\i}ld{\i}z, O.}
\newblock \bibinfo{journal}{\bibinfo{title}{The effect of chronic diseases, age
  and gender on morbidity and mortality of {COVID-19} infection}}.
\newblock {\emph{\JournalTitle{Iran. J. Public Health}}}
  \textbf{\bibinfo{volume}{50}}, \bibinfo{pages}{721--727}
  (\bibinfo{year}{2021}).

\bibitem{10.1093/infdis/jiac063}
\bibinfo{author}{Nyberg, T.} \emph{et~al.}
\newblock \bibinfo{journal}{\bibinfo{title}{{Hospitalization and Mortality Risk
  for COVID-19 Cases With SARS-CoV-2 AY.4.2 (VUI-21OCT-01) Compared to
  Non-AY.4.2 Delta Variant Sublineages }}}.
\newblock {\emph{\JournalTitle{The Journal of Infectious Diseases}}}
  \textbf{\bibinfo{volume}{226}}, \bibinfo{pages}{808--811},
  \doiprefix\url{10.1093/infdis/jiac063} (\bibinfo{year}{2022}).
\newblock
  \eprint{https://academic.oup.com/jid/article-pdf/226/5/808/45819803/jiac063.pdf}.

\end{thebibliography}
\newpage
\section*{Supplementary Materials}

\renewcommand{\thefigure}{S\arabic{figure}}
\setcounter{figure}{0}

\subsection*{Risk of infection in the case of Delhi, India}
Our main focus in carrying out this study was to create an easy-to-use risk calculator that could be implemented for a large set of countries with the least amount of internet bandwidth needed. To test out whether the data we extracted as well as the workflow we followed was effective in accurately estimating risks associated with carrying out a daily life activity for an individual. We carried out a secondary test taking the example of Delhi, India.

\begin{figure}[hbt!]
\centering
\includegraphics[width=\linewidth]{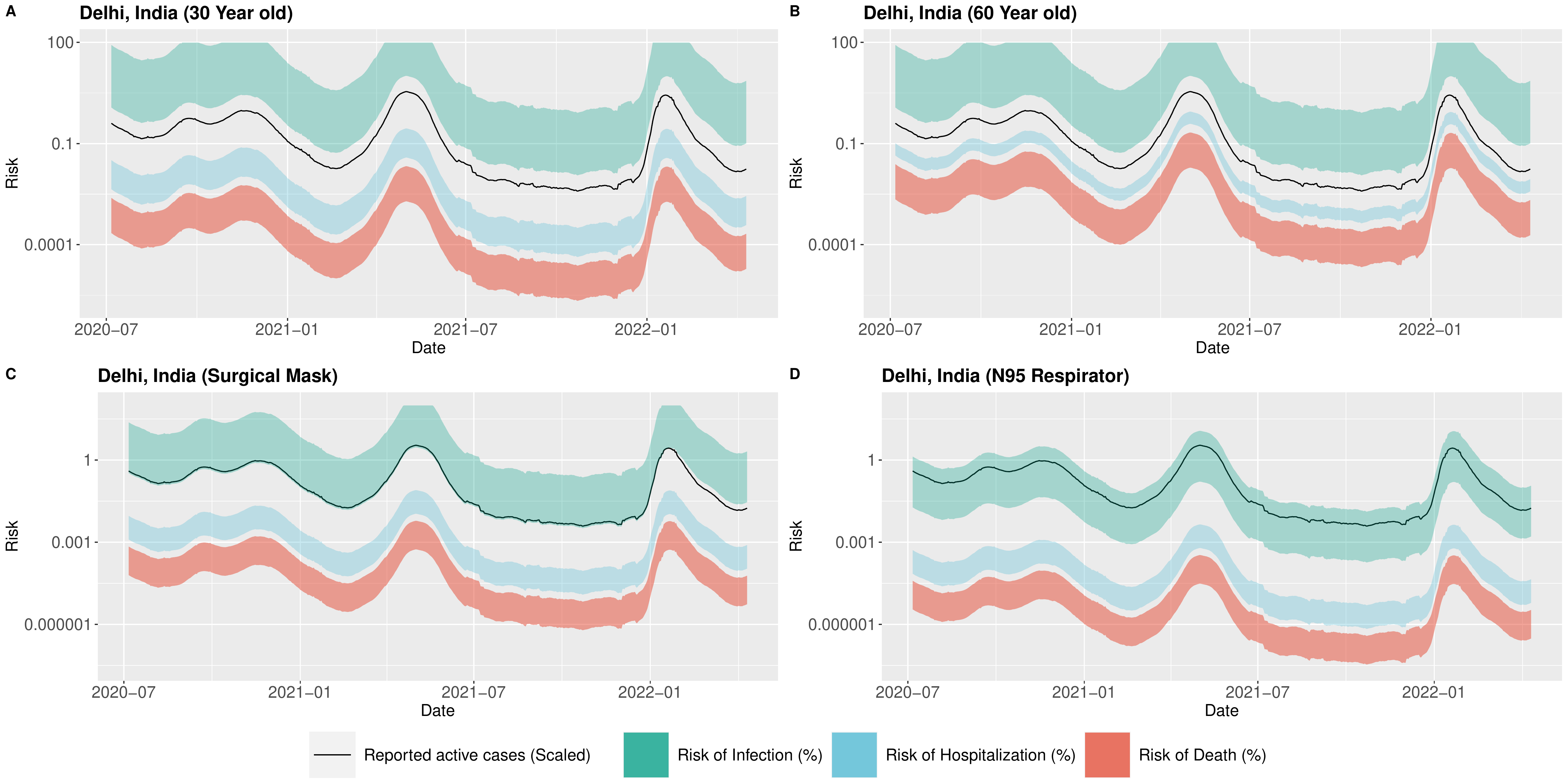}
\includegraphics[width=\linewidth]{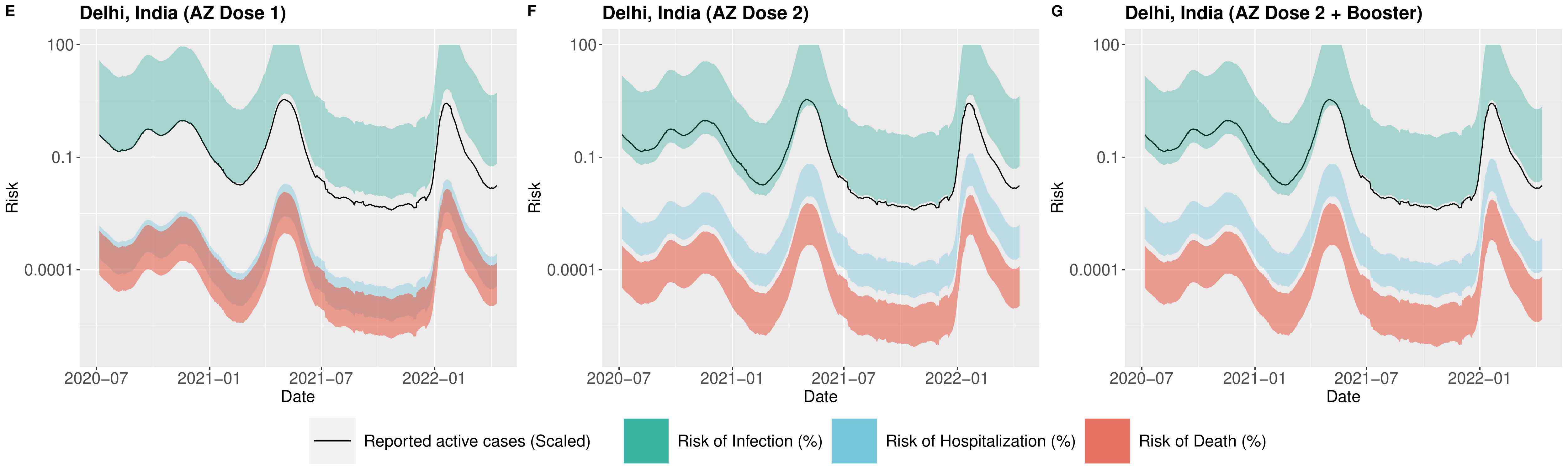}
\caption{\small{In the following figures, we use the location of Delhi, India and calculated the ranges of risk of infection, hospitalization and death for a \textbf{A.} 30-year-old male with no chronic illness, no mask and no vaccination, 10 people passed outdoors and 5 people passed indoors during the activity, a \textbf{B.} 60-year-old male with no chronic illness, no mask and no vaccination, 10 people passed outdoors and 5 people passed indoors during the activity. a \textbf{C.} 30-year-old male with no chronic illness, surgical mask and no vaccination when 10 people are passed outdoors and 5 people passed indoors during the activity, a \textbf{D.} 30-year-old male with no chronic illness, N95 respirator mask and no vaccination when 10 people are passed outdoors and 5 people passed indoors during the activity, a \textbf{E.} 30 year old male with no chronic illness, no mask and Dose 1 of AstraZeneca vaccination when 10 people are passed outdoors and 5 people indoors during the activity, a \textbf{F.} 30-year-old male with no chronic illness, no mask and Dose 2 of AstraZeneca vaccination when 10 people are passed outdoors and 5 people indoors during the activity and a \textbf{G.} 30-year-old male with no chronic illness, no mask and Dose 2 with a booster dose of AstraZeneca vaccination when 10 people are passed outdoors and 5 people indoors during the activity.}}
\label{fig:graph_2}
\end{figure}

In order to further validate the robustness of our risk calculator in the estimation of risk associated with COVID-19 in different countries, we carried out a secondary analysis where we used the location of Delhi, India, in order to estimate different risks associated with COVID-19 when carrying out a daily life activity. Using our risk calculator, we performed a test similar to the previously performed risk estimation for Franklin, MA, USA but for the case of Delhi, India. We did not have county-level information therefore, we carried out the test on a city level. We first calculated the 
 range of risk of infection, hospitalization, and death with change time for a 30-year-old and 70-year-old male living in Delhi, India, with no chronic illness, no vaccination, no mask, five people passed indoors, and ten people passed outdoors (see \href{fig:graph_2}{Supplementary Figure \ref{fig:graph_2} A., B.}). We then conducted the same test but only took the case of a 30-year-old male when AstraZeneca vaccination first, second and booster doses were taken in order to check the reduced range of risk of infection. (see \href{fig:graph_2}{Supplementary Figure \ref{fig:graph_2} E., F., G.}). We then also carried out the tests related to the reduction in range of risk of infection when different types of masks are used (see \href{fig:graph_2}{Supplementary Figure \ref{fig:graph_2} C., D.}). 

As we can see with our results, we were able to observe an increase in the risk of hospitalization and risk of death with an increase in the age of the user. Furthermore, there is a high risk of hospitalization and death during the second wave (when the delta variant was observed) compared to the third wave (when the omicron variant was observed). There is also a reduction in risk of infection when different doses of vaccination are taken by the user which is similar to what we observed in the case of Franklin, MA, USA. Furthermore, the reduction in risk is also very high when different types of masks are used by the user which is analogous to the results that we previously obtained. 

\subsection*{Calculation of Risk for different Countries}

\begin{figure}[hbt!]
\centering
\includegraphics[width=\linewidth]{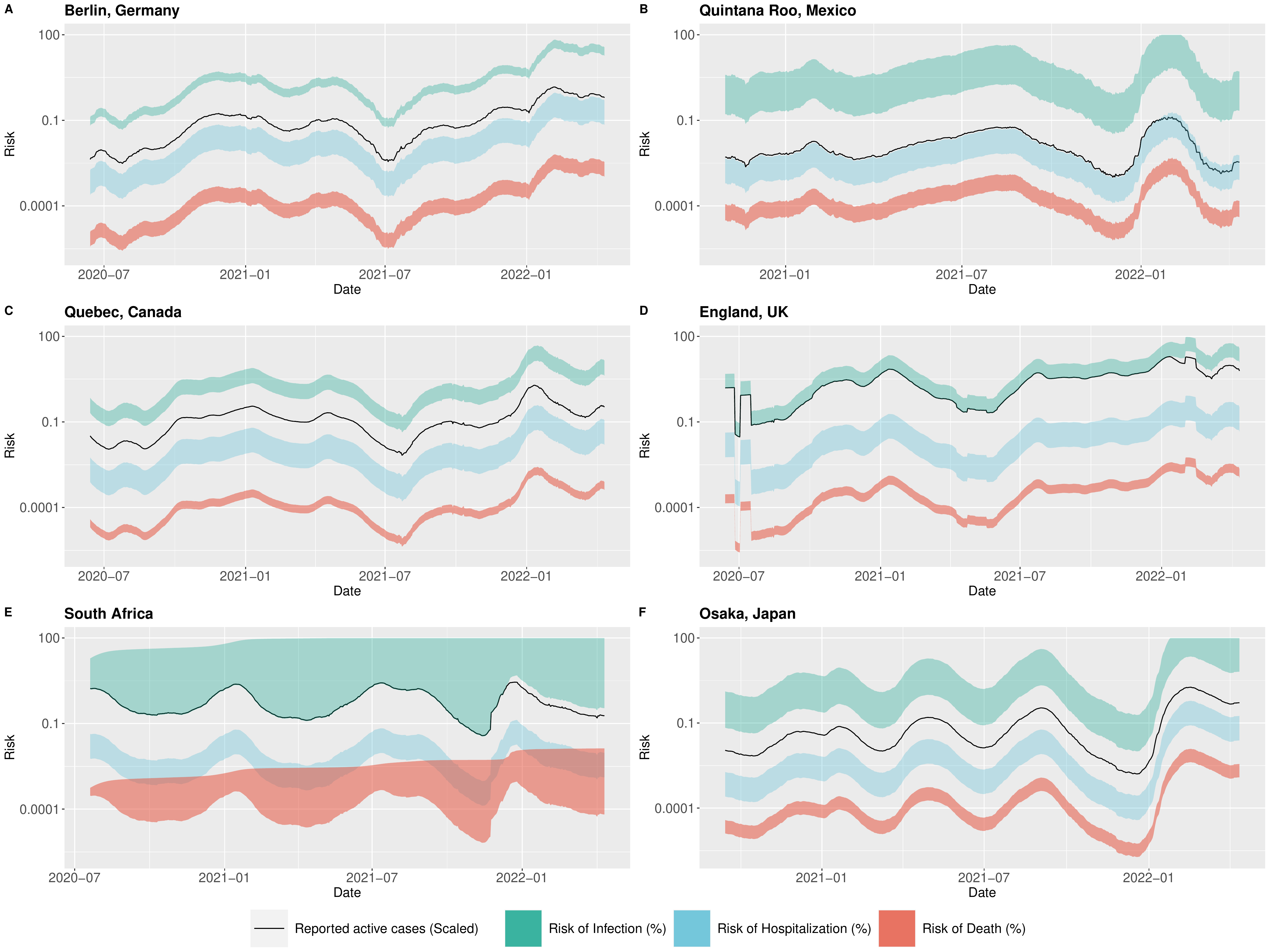}
\caption{\small{In the given figure, we estimated the ranges of risk of infection, hospitalization, and death for a 30-year-old male with no past chronic illness, no mask, and no vaccination carrying out an activity involving interaction with 10 people outdoors and 5 people indoors in the locations \textbf{A.} Berlin (Germany), \textbf{B.} Quintana Roo (Mexico), \textbf{C.} Quebec (Canada), \textbf{D.} England, United Kingdom, \textbf{E.} South Africa and \textbf{F.} Osaka, Japan}}
\label{fig:graph_country}
\end{figure}

We further tested out the risk calculator over the data for 6 other regions and countries. The regions included Berlin (Germany), Quintana Roo (Mexico), Quebec (Canada), England (United Kingdom), Osaka (Japan) and South Africa. Since South Africa did not provide regional-level information, we estimated the risk on a country level (see \href{fig:graph_2}{Supplementary Figure \ref{fig:graph_country} A., B., C., D., E., F.}). 

For each of the regions, we estimated the risk by taking an example of a 30-year-old male who is going to carry out an activity involving interaction with 10 people outdoors and 5 people indoors, with no past chronic illness, no masks and no vaccinations. The results were consistent with our previous findings. The risk scores varied across regions, with some regions experiencing higher risk levels than others. For instance, Berlin (Germany) had a relatively low-risk score compared to South Africa or Quintana Roo (Mexico), which had a significantly higher risk score. Overall, the results demonstrated the effectiveness of the COVID-19 risk calculator in assessing the risk levels of different regions and countries. 

Using this test we were able to clearly check the validity of our risk calculator for different regions and countries across the world. It helped us in ensuring that the calculator is useful for estimating several individual-level risks related to carrying out any specific activity during COVID-19 for 203 countries across the world whose data we also had available.

\subsection*{Calculation of Risk for different Age Groups}

\begin{figure}[hbt!]
\centering
\includegraphics[width=\linewidth]{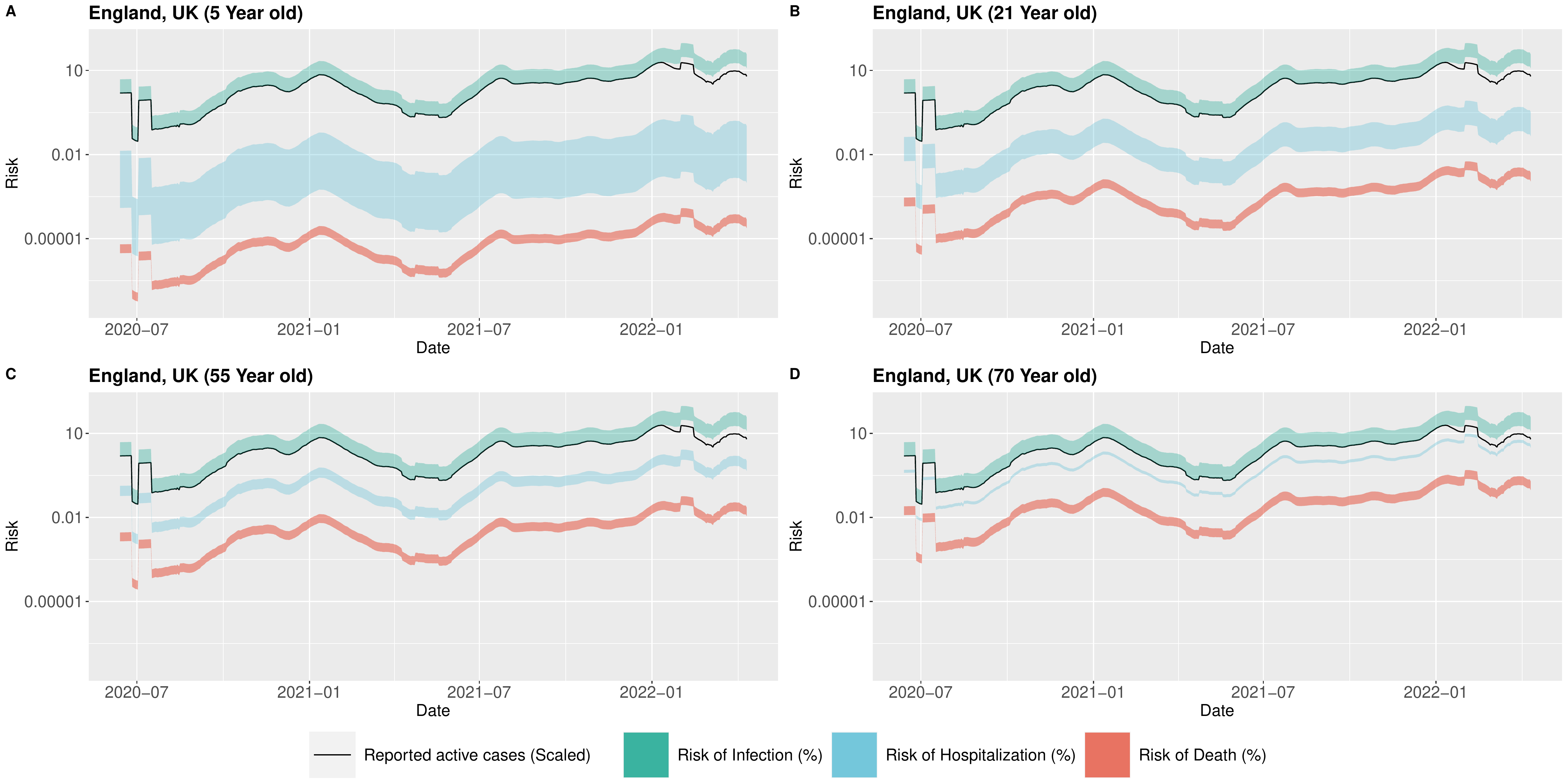}
\caption{\small{In the given figure, we estimated the ranges of risk of infection, hospitalization, and death for a male with no past chronic illness, no mask, and no vaccination carrying out an activity involving interaction with 10 people outdoors and 5 people indoors in the location of England (United Kingdom) at the age of \textbf{A.} 5 years, \textbf{B.} 21 years (M, \textbf{C.} 55 years and \textbf{D.} 71 years.}}
\label{fig:graph_age}
\end{figure}

The next test that we wanted to perform was to estimate the change in the risk of infection, hospitalization and death across different age groups. For this case, we used the example of England (United Kingdom) and estimated change in the trend of risk of infection, hospitalization and death for a 5-year-old, 21-year-old, 55-year-old and 70-year-old male with no past chronic illness, no mask and no vaccination performing an activity involving interaction with 10 people outdoors and 5 people indoors (see \href{fig:graph_age}{Supplementary Figure \ref{fig:graph_age} A., B., C., D.}).

We found that the risk of infection, hospitalization, and death varies significantly across age groups, with older individuals having a higher risk compared to younger individuals. For instance, for the example scenario in England, the risk of infection was found to be the highest for the 70-year-old male, followed by the 55-year-old, 21-year-old, and 5-year-old males, in that order. Similarly, the risk of hospitalization and death also increased with age. These findings highlight the importance of age as a significant factor in assessing COVID-19 risk and the need for age-specific risk assessment tools. This inspired us to include age-specific risk estimates in future version updates of CovARC to further improve its accuracy and usefulness.

\subsection*{Simulating different scenarios for variants}
\begin{figure}[htb!]
\centering
\includegraphics[width=\linewidth]{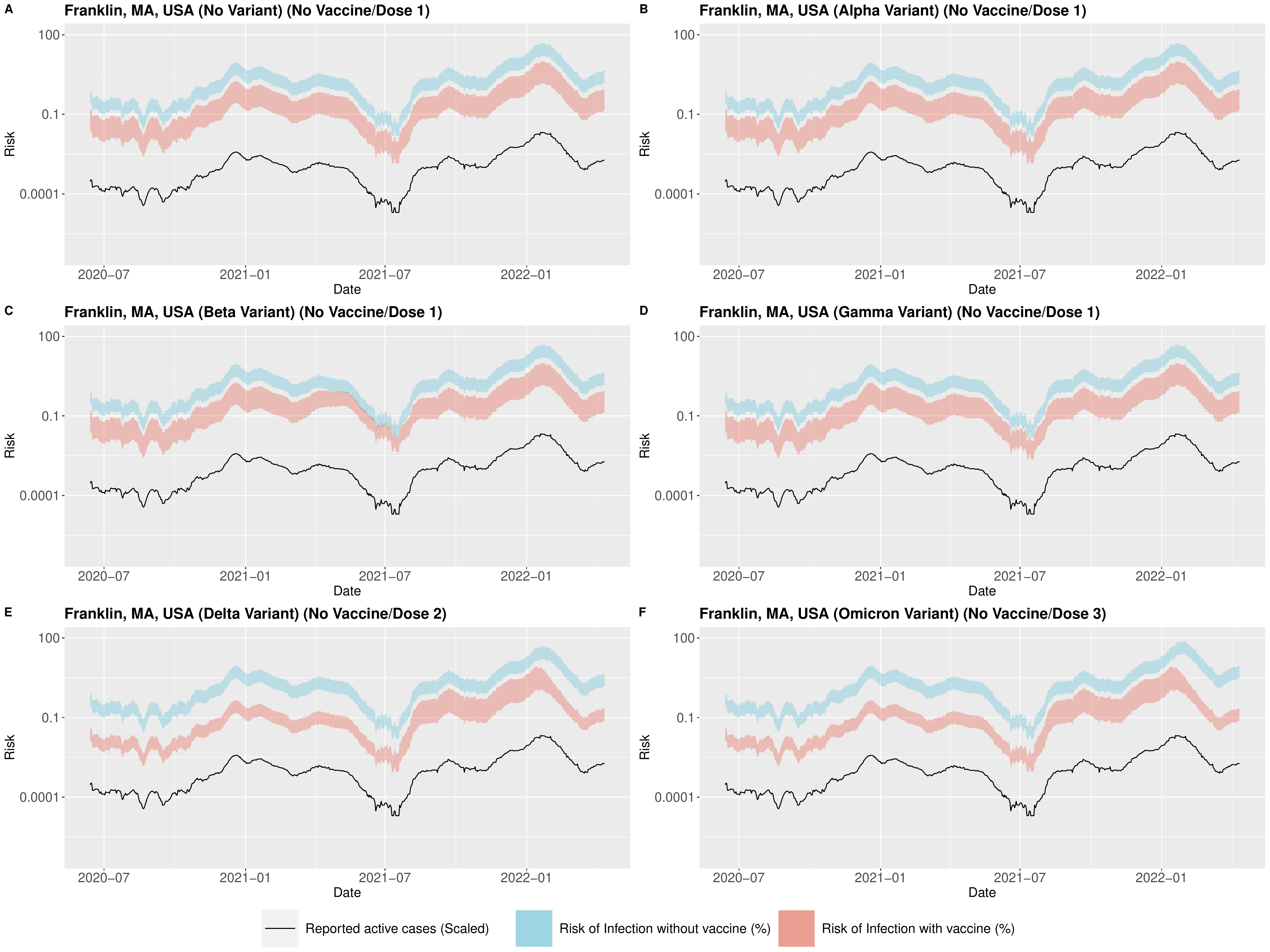}
\caption{\small{The given figure illustrates the ranges of risk of infection for a simulated scenario of a 30-year-old male with no past chronic illness and no mask vaccination carrying out an activity in Franklin, Massachusetts, USA involving interaction with 10 people outdoors and 5 people indoors when there is the presence of \textbf{A.} No variants with no vaccination and no variants with Dose 1 of Pfizer vaccination, \textbf{B.} Only Alpha variant with no vaccination and only Alpha variant with Dose 1 of Pfizer Vaccination, \textbf{C.} Only Beta variant and no vaccination and only Beta variant with Dose 1 of Pfizer Vaccination, \textbf{D.} Only Gamma variant with no vaccination and only Gamma variant with Dose 1 of Pfizer Vaccination, \textbf{E.} Only Delta variant and no vaccination and only Delta variant with Dose 2 of Pfizer Vaccination, \textbf{E.} Only Omicron variant and no vaccination and only Omicron variant with Dose 2 \& Booster dose of Pfizer Vaccination}}
\label{fig:graph_variants}
\end{figure}

As the next step, we wanted to simulate different scenarios where there was the presence of either no variant or one specific variant to check the change in the trend of ranges of risk of infection when a specific dosage of vaccination was taken and when no vaccination was taken. In order to do so, we created 6 scenarios wherein there was the presence of no variant, only alpha variant, only beta variant, only gamma variant, only delta variant and only omicron variant. We then simulated a scenario where there was Dose 1 of Pfizer vaccination taken for the case of no variant presence and only alpha or beta or gamma variant presence. Furthermore, we simulated another case where Dose 2 of the Pfizer vaccination was taken in the presence of only the delta variant and a last case where Dose 2 and a booster dose of the Pfizer vaccination were taken in the presence of only the Omicron variant. The test was performed for a 30-year-old male living in Franklin, Massachusetts, USA with no past chronic illness and mask carrying out an activity involving interaction with 10 people outdoors and 5 people indoors (see \href{fig:graph_variants}{Supplementary Figure \ref{fig:graph_variants} A., B., C., D., E., F.}).

In addition to the effectiveness of different vaccine doses for different variants, the simulation also allowed us to observe the impact of variant presence on the ranges of risk of infection. The results demonstrated that the range of risk of infection was significantly higher in the presence of the delta and omicron variants compared to the other variants and no variant scenario. This highlights the importance of continued monitoring and research on the impact of emerging variants on COVID-19 transmission and the efficacy of vaccines. The simulation also provided valuable insights into the potential benefits of booster doses for the omicron variants, which could inform future vaccination strategies. Overall, the results of this simulation emphasize the importance of taking into account the presence of variants when evaluating the effectiveness of vaccines and assessing the ranges of risk of infection.

\end{document}